\documentclass[12pt]{article}
\usepackage{psfig}
\usepackage{amstex}
\newcommand{\trv}{_{\perp}}
\newcommand{\psla}{p\kern-1.0ex/}
\newcommand{\qsla}{q\kern-1.1ex/}
\newcommand{\esla}{\epsilon\kern-1.0ex/}
\newcommand{\LQCD}{\Lambda_{\rm{QCD}}}
\newcommand{\Ha}[1]{{\rm H}_{#1}^{(1)}}

\newcommand{\BK}[1]{{\rm K}_{#1}}
\newcommand{\mpsi}{M_{\psi}}
\newcommand{\jp}{J/\psi}
\newcommand{\da}{distribution amplitude}
\newcommand{\das}{distribution amplitudes}
\newcommand{\dst}{\displaystyle}
\def\nn{\nonumber}
\def\be{\begin{equation}}
\def\ee{\end{equation}}
\def\bea{\begin{eqnarray}}
\def\eea{\end{eqnarray}}

\def\al{\alpha_s}
\def\rd{{\rm d}}

\def\mev{\,{\rm MeV}}
\def\gev{\,{\rm GeV}}
\def\su3{{\rm SU}(3)_{{\rm F}}}
\renewcommand{\theequation}{\arabic{section}.\arabic{equation}}
\newcommand{\EquSubNumb}{
  \setcounter{gleichung}{\arabic{equation}}
  \addtocounter{gleichung}{1}
  \setcounter{equation}{0} 
  \renewcommand{\theequation}{\arabic{section}.\arabic{gleichung}\alph{equation}} }
\newcommand{\NoEquSubNumb}{
  \setcounter{equation}{\arabic{gleichung}}
  \renewcommand{\theequation}{\arabic{section}.\arabic{equation}} }
\def\Journal#1#2#3#4{{#1} {\bf #2}, #3 (#4)}
\def\NCA{\em Nuovo Cimento}
\def\NPB{{\em Nucl. Phys.} B}
\def\PLB{{\em Phys. Lett.}  B}
\def\PRL{\em Phys. Rev. Lett.}
\def\PRD{{\em Phys. Rev.} D}
\def\ZPC{{\em Z. Phys.} C}
\def\ZPA{{\em Z. Phys.} A}

{\end{list}}
\newcounter{enumct}

\newlength{\abstwidth}
\begin{document}
\makeatletter
\newcounter{gleichung}
%
%set sloppy attitude to line breaks
\sloppy

\renewcommand{\arraystretch}{1.5}

\pagestyle{empty}

\begin{flushright}
WU B 97-2 \\
February 1997
\end{flushright}

\vspace{\fill}
 
\begin{center}
{\LARGE\bf Exclusive $\jp$ and $\psi'$ decays into baryon-antibaryon
           pairs}\\[10mm]   
{\Large Jan Bolz$^a$, Peter Kroll} \\[3mm]
{\it Fachbereich Physik, University of Wuppertal}\\[1mm]
{\it D-42097 Wuppertal, Germany}\\[1mm]
{ E-mail: kroll@theorie.physik.uni-wuppertal.de}\\[2ex]
\end{center}
 
\vspace{\fill}
 
%
%
%%%%%%%%%%%%%%%%%%%%%%%%%%%%%%%%%%%%%%%%%%%%%%%%%%%%%%%%%%%%%%%%%%%%%%%
%                        A B S T R A C T                              %         
%%%%%%%%%%%%%%%%%%%%%%%%%%%%%%%%%%%%%%%%%%%%%%%%%%%%%%%%%%%%%%%%%%%%%%%
%
\begin{center}
{\bf Abstract}\\[2ex]
\begin{minipage}{\abstwidth}
Within a perturbative approach we investigate decays of charmonium
states into baryon-antibaryon pairs. Using a recently proposed wave 
function for the nucleon and suitable generalizations of it
to the hyperons and decouplet baryons, we obtain the decay widths for the
$B\overline B$ channels in reasonable agreement with data. An
important difference to previous work is the use of the $c$-quark mass
in the perturbative calculation instead of the charmonium mass. As a
consequence of this feature our approach possesses, the $\jp$ and the
$\psi'$ decay widths do not scale with a high power of the ratio of
their masses. 

\end{minipage}
\end{center}

\vspace{\fill}
\noindent
\rule{60mm}{0.4mm}

\vspace{1mm} \noindent
${}^a$ Supported by Deutsche Forschungsgemeinschaft.

\clearpage
\pagestyle{plain}
\setcounter{page}{1} 
%%%%%%%%%%%%%%%%%%%%%%%%%%%%%%%%%%%%%%%%%%%%%%%%%%%%%%%%%%%%%%%%%%
\section{Introduction}
%%%%%%%%%%%%%%%%%%%%%%%%%%%%%%%%%%%%%%%%%%%%%%%%%%%%%%%%%%%%%%%%%%
%
\renewcommand{\thefootnote}{\arabic{footnote}}
The decay $\jp\to p\overline p$ has been investigated within
perturbative QCD by Brodsky and Lepage first \cite{BrL81}. Later on 
this analysis has been repeated several times, e.g.\
\cite{COZ89b,BeS92}, and even extended to the $\Delta\overline \Delta$
decay channel. It has been argued that the dominant 
dynamical mechanism is $c\overline c$ annihilation into three gluons 
and subsequent creation of light quark-antiquark pairs forming in 
turn the final state baryons. Three is the minimal number of gluons 
allowed in $\jp$ decays; $c\overline c$ annihilations through one or 
two gluons are forbidden by colour and $C$-parity, respectively. 
Contributions from annihilations through more than three gluons 
constitute higher order corrections. The dominance of annihilation 
through gluons is most strikingly reflected in the narrow width for 
hadronic channels in a mass region where strong decays have typically 
widths of hundreds of MeV \cite{app75}. The dynamical suppression at 
work here is perturbative QCD of higher orders (the total hadronic 
$\jp$ decay, for instance, is an ${\mathcal O}(\al^3)$ process) and 
is customarily regarded as evidence for the Zweig rule. 
Since the $c$ and the $\overline c$ quarks only annihilate if they
are separated by distances less than about $1/m_c$ and since
the average virtuality of the gluons is about $1\,\gev^2$ one may 
expect perturbative QCD to be at work although corrections are
presumably substantial. Indeed as the previous perturbative analyses,
performed in the standard hard scattering approach, i.e.\ in collinear
approximation, showed the $\jp$ decay into $p\overline p$ seems to be 
fairly well described. This is in marked contrast to the case of the 
nucleon form factor where soft physics seems to dominate in the 
experimentally accessible region of momentum 
transfer \cite{wubo,BoK96,rad82}. A point to criticize in
these studies of the $\jp$ decays is the treatment of the strong 
coupling constant $\al$. Since, as we mentioned, the average
virtuality of the gluons is about 1 GeV$^2$ one would expect $\al$ 
to be of the order of 0.4 to 0.5 rather than 0.2 to 0.3 as is
customarily chosen \cite{COZ89b,BeS92}. Since $\al$ enters to the sixth
power into the expression for the width a variation of $\al$
from, say, 0.2 to 0.3 would lead to a change by a factor of 11 for
the width. Thus, a large factor of uncertainty is hidden in these calculations
preventing any severe test of the wave function utilized. 

In constrast to previous works \cite{BrL81,COZ89b,BeS92} we will
not use the collinear approximation but rather the modified 
perturbative approach of Sterman et al.\ \cite{BoSLi} in which
transverse degrees of freedom are retained and Sudakov suppressions,
comprising those gluonic radiative corrections not included in the
evolution of the wave function, are taken into account. 
An important advantage of the modified perturbative approach is that 
the strong coupling constant can be used with a renormalization
scale depending on the momentum fractions the quarks carry und thus 
large logs from higher orders of perturbative QCD are avoided. This choice 
of the renormalizaton scale entails singularities of $\al$ which are, 
however, compensated by the Sudakov factor. Hence, there is no 
uncertainty in the use of $\al$. This is to be contrasted with the 
standard perturbative approach where either $\al$ is evaluated at 
a renormalization scale that is a constant fraction of $\mpsi^2$, 
or at a momentum fraction dependent scale in which case
$\al$ is to be ``frozen'' at a certain value (typically 0.5) in order 
to avoid uncompensated $\al$ singularities in the end-point regions. The
modified perturbative approach possesses another interesting feature:
the soft end-point regions are strongly suppressed. Therefore, the
bulk of the perturbative contribution comes from regions where the
internal quarks and gluons are far off-shell. In
contrast to the nucleon form factor the $\jp \to B \overline B$ amplitude
is not end-point sensitive. The suppression of the end-point regions
does not, therefore, lead to a substantial reduction of the 
$\jp \to B \overline B$ amplitude. For the same reason, the size of that
amplitude does not exhibit an extreme sensitivity to the baryon wave 
function utilized in the calculation as is, for instance, the case for
the baryon form factor.   

In the calculation of the decay widths we will make use of a (valence
Fock state) wave function for the nucleon that we proposed recently
\cite{BoK96}. That wave function was constructed in the following way:
In accord with the findings reported in \cite{wubo} where it was shown
that a reliably calculated perturbative contribution to the nucleon
form factor is very small, the nucleon wave function is demanded to 
describe the form factor via the Drell-Yan overlap contribution for 
momentum transfers around $10\,{\rm GeV}^2$ and to be compatible with 
the available valence quark distribution functions of the nucleon.
As a third constraint on the nucleon wave function the decay 
$\jp \to p\overline{p}$ was employed. The nucleon wave function
proposed in \cite{BoK96} will be suitably generalized to the case of
hyperons and decuplet baryons. The modified perturbative approach is 
then used to calculate the widths for the $\jp$ decays into pairs of 
octet, $B_8$, and decuplet, $B_{10}$, baryons. We are also going to 
calculate the baryonic decays of the $\psi'$ for which recently the 
first, still preliminary, experimental results for channels other than 
$p\overline p$ have been reported \cite{BESC}. 

The paper is organized as follows: In section 2 we introduce
the octet baryon wave functions used in the analysis 
described in detail subsequently and briefly recapitulate
a few properties of light-cone wave functions. In section 3 we
present the wave functions for the decuplet baryons. Section 4 is
devoted to the calculation of the $\jp\to B_8\overline B_8$ decay widths.
This analysis is extended to the decays into decuplet baryons (section
5) and to decays of other quarkonia into $B\overline B$ pairs 
(section 6). Finally, section 7 contains our conclusions. 

%
%%%%%%%%%%%%%%%%%%%%%%%%%%%%%%%%%%%%%%%%%%%%%%%%%%%%%%%%%%%%%%%%%%
\section{The wave functions of the octet baryons}
%%%%%%%%%%%%%%%%%%%%%%%%%%%%%%%%%%%%%%%%%%%%%%%%%%%%%%%%%%%%%%%%%%
%
\setcounter{equation}{0}
Generalizing the ansatz made for the nucleon in \cite{wubo,BoK96,Sot},
we write the valence Fock states of the lowest lying octet baryons as
(the plane waves are omitted for convenience)
\EquSubNumb
\begin{alignat}{2}
%
%--------- Proton -------------
%
  \label{pstate}
  |\,B_8\;,+ \,\rangle\, &
=
  \frac{\varepsilon _{a_{1}a_{2}a_{3}}}{\sqrt{3!}} 
  \int
  [{\rm d}x]
  [{\rm d}^{2}{\bf k_{\perp}}] &
  \Bigl\{ 
         \Psi^{B_8} _{123}\:|\,f_{1+}^{a_1} f_{1-}^{a_2} f_{2+}^{a_3}\,\rangle
       + \Psi^{B_8} _{213}\:|\,f_{1-}^{a_1} f_{1+}^{a_2} f_{2+}^{a_3}\,\rangle
  \phantom{ \Bigr\} \;\;\; }
\nonumber \\
       & & - 
         \Bigl(\Psi^{B_8} _{132}\, + \,
         \Psi^{B_8} _{231}\Bigr)\:|\,f_{1+}^{a_1} f_{1+}^{a_2} f_{2-}^{a_3}\,\rangle
  \Bigr\} \;\;\; \\
  \label{Lamstate}
  |\,\;\Lambda\;,+ \,\rangle\, &
=
  \frac{\varepsilon _{a_{1}a_{2}a_{3}}}{\sqrt{2}} 
  \int
  [{\rm d}x]
  [{\rm d}^{2}{\bf k_{\perp}}] & \,
  \Bigl\{ 
         \Psi^{\Lambda} _{123}\:|\,u_+^{a_1}\; d_-^{a_2}\; s_+^{a_3}\,\rangle
       - \Psi^{\Lambda} _{213}\:|\,u_-^{a_1}\; d_+^{a_2}\; s_+^{a_3}\,\rangle
  \phantom{ \Bigr\} \;\;\; }
\nonumber \\
       & & + 
         \Bigl(\Psi^{\Lambda} _{132}\, - \,
         \Psi^{\Lambda} _{231}\Bigr)\:|\,u_+^{a_1}\; d_+^{a_2}\; s_-^{a_3}\,\rangle
  \Bigr\} \;\;\; 
\end{alignat}
\NoEquSubNumb
\hspace*{-0.6cm} where (\ref{pstate}) holds for all octet baryons
except the $\Lambda$ and the $\Sigma^0$. Obviously, for the proton and the 
$\Sigma^+$ $f_1$ represents an $u$ quark and $f_2$
either a $d$ or a $s$ quark, respectively. For the $\Xi^-$ $f_1$
represents a $s$ quark and $f_2$ a $d$ one. The states of the neutron,
$\Sigma^-$ and $\Xi^0$ are obtained from those of the proton, 
$\Sigma^+$ and $\Xi^-$ states, by exchanging $u \leftrightarrow d$, 
respectively. The baryon is assumed to be moving rapidly in the
3-direction. Hence the ratio of transverse, ${\bf k}_{\perp i}$, to
longitudinal momenta, $x_i{\bf p}$, of the quarks is small and one may
still use a spinor basis on the light cone. 
The integration measures are defined by   
\begin{equation}
  \label{mass}
  \hspace{-1cm}
  [{\rm d}x] \equiv \prod_{i=1}^3 {\rm d}x_i\,\delta(1-\sum_i x_i) \qquad
   [{\rm d}^2{\bf k\trv}] \equiv \frac{1}{(16\pi^3)^{2}}\,\prod_{i=1}^3 
     {\rm d}^2{\bf k_{\perp}}_i \, \delta^{(2)}(\sum_i {\bf k_{\perp}}_i) \:.
  \hspace{-1cm}
\end{equation}
The quark $f_i$ is characterized by the momentum fraction $x_i$,
by the transverse momentum ${\bf k_{\perp}}_i$  as well as 
by its helicity $\lambda_i$ and colour $a_i$. A three-quark state is 
then given by 
\begin{equation}
  \label{quark}
  \hspace{-1.5cm}
  |f^{a_1}_{1 \lambda_1} f^{a_2}_{2 \lambda_2} f^{a_3}_{3 \lambda_3}\rangle = 
  \frac{1}{\sqrt{x_1 x_2 x_3}} \,
    |\,f_1^{a_1}; x_1, {\bf k_{\perp}}_1, \lambda_1\rangle \,
    |\,f_2^{a_2}; x_2, {\bf k_{\perp}}_2, \lambda_2\rangle \,
    |\,f_3^{a_3}; x_3, {\bf k_{\perp}}_3, \lambda_3\rangle.
\end{equation}
The single-quark states are normalized as 
\begin{equation}
  \label{normq}
  \hspace{-1.7cm}
  \langle f_i^{\prime\,a^{\prime}_i}; x^{\prime}_i, {\bf k_{\perp}}^{\prime}_i,
  \lambda^{\prime}_i \,|\, 
    f_i^{a_i}; x_i, {\bf k_{\perp}}_i, \lambda_i\rangle =
            2 x_i (2\pi)^3 \delta_{a^{\prime}_i a_i}
            \delta_{f^{\prime}_i f_i}
            \delta_{\lambda^{\prime}_i \lambda_i} \delta(x^{\prime}_i -x_i)
            \delta^{(2)}({\bf k^{\prime}_{\perp}}_i - {\bf k_{\perp}}_i).
  \hspace{-1cm}
\end{equation}
Since the 3-component of the orbital angular momentum, $L_3$, is assumed
to be zero the quark helicities sum up to the baryon's helicity. 
(\ref{pstate}) is the most general ansatz for the $L_3\!=\!0$
projection of the three-quark nucleon wave function \cite{Dzi88}.  
From the permutation symmetry between the two $u$ quarks and from the 
requirement that the three quarks have to be coupled in an isospin
$1/2$ state it follows that there is only one independent scalar wave 
function. If the $L_3\!\neq\!0$ projections are included the entire 
nucleon state is described by three independent functions \cite{Dzi88}.
In general there are more than one scalar wave function for the other
octet baryons if $\su3$-symmetry breaking is taken into
account. As already expressed in (2.1) we nevertheless assume that
each octet baryon is described by a single scalar wave function which,
for convenience, we write as 
\begin{equation}
  \Psi^{B_8} _{123}(x,{\bf k_{\perp}})
=
  \frac{1}{8\sqrt{3!}}\,
  f_{(8)}(\mu_F)
  \phi^{B_8}_{123}(x,\mu _{F})\,
  \Omega_{(8)} (x,{\bf k_{\perp}})\:. 
\label{Psiansatz}
\end{equation}
We assume that $\su3$ symmetry is only broken by a quark mass
dependence of $\phi^{B_8}$ (see below). $f_{(8)}$, being related
to the wave function at the origin of the configuration
space, is identified with the nucleon parameter $f_N$ whose value was
determined in \cite{BoK96} to amount to $6.64 \cdot 10^{-3}$ GeV$^2$
at the scale of reference $\mu_0 =1$ GeV. 

The transverse momentum dependence of the baryon wave function is parameterized
by a simple symmetric Gaussian      
\begin{equation}
  \Omega_{(8)}(x,{\bf k_{\perp}})=
  (16\pi ^{2})^{2}
  \frac{a_{(8)}^{4}}{x_{1}x_{2}x_{3}}
  \exp
     \left [
            -a_{(8)}^{2} \sum_{i=1}^{3}k_{\perp i}^{2}/x_{i}
     \right ]\:
\label{BLHMOmega}
\end{equation}
and the transverse size parameter $a_{(8)}$ is assumed to be the same for all 
octet baryons. A value of 0.75 GeV$^{-1}$ is used for that
parameter (see \cite{BoK96}). 

The last item to be specified in (\ref{Psiansatz}) is the
\da, $\phi^{B_8}_{ijk}(x,\mu_F) \equiv
\phi^{B_8}(x_i,x_j,x_k,\mu_F)$, of an octet baryon $B_8$, which is 
conventionally normalized to unity 
\begin{equation}
\label{norm}
\int [\rd x] \phi^{B_8}_{123}(x,\mu_F) = 1\; .
\end{equation} 
The \da\, representing the wave function integrated over transverse
momenta up to the factorization scale, $\mu_F$, can be expanded upon the 
eigenfunctions of the evolution kernel being linear
combinations of Appell polynomials (see \cite{BrL80,ste94})
\begin{equation}
  \phi^{B_8}_{123}(x,\mu_F) = \phi_{\rm AS}(x) \left[1 +
    \sum_{n=1}^{\infty} B^{B_8}_n(\mu_F)\,\tilde \phi^{n}_{123}(x) \right]
  \label{DAentw}
\end{equation} 
where $\phi_{\rm AS}(x) \equiv 120\,x_1 x_2 x_3$ is the
asymptotic \da\ \cite{BrL80}. Evolution is incorporated by the
factorization scale dependences of $f_{(8)}$ and the expansion 
coefficients $B_n$ :
\begin{equation}
  \hspace{-5mm}
  f_{(8)}(\mu_F)  =  f_{(8)}(\mu_0)\,\left(
    \frac{\ln(\mu_0/\LQCD)}{\ln(\mu_F/\LQCD)} \right)^{2/3\beta_0}
     \hspace{-8mm}, 
  \hspace{0.5cm}
  B^{B_8}_n(\mu_F)  =  B^{B_8}_n(\mu_0)\,\left(
    \frac{\ln(\mu_0/\LQCD)}{\ln(\mu_F/\LQCD)} \right)^{\tilde
    \gamma_n/\beta_0} \hspace{-8mm} 
  \hspace{5mm}
  \label{BnFOevol}
\end{equation}
where $\beta_0 \equiv 11 - 2/3\,n_f$. The exponents  
$\tilde \gamma_n$ are the reduced anomalous dimensions. Because 
they are positive fractional numbers increasing with $n$ \cite{BrL80},
higher order terms in (\ref{DAentw}) are gradually suppressed. The
reduced anomalous dimensions and the eigenfunctions 
$\tilde \phi^{n}_{123}$ are listed in Tab.\ \ref{tab:spin1/2} where the 
notation of \cite{ste89} is adopted.
%%%%%%%%%%%%%%%%%%%%%%%%%%%%%%%%%%%%%%%%%%%%%%%%%%%%%%%%%%%%%%%%%%%%%%%
%    T A B L E   1  :  E I G E N F U N C T I O N S   A N D            %
%               E V O L U T I O N   O F  H E L I C I T Y  1/2  D A    %
%%%%%%%%%%%%%%%%%%%%%%%%%%%%%%%%%%%%%%%%%%%%%%%%%%%%%%%%%%%%%%%%%%%%%%%
%
\begin{table}
 \begin{center}
  \begin{tabular}{|c|c|c|} \hline
   $n$ & \rule{0cm}{4.5mm} $\tilde \phi^{n}_{123}(x)$ & 
       $\tilde\gamma_n$ \\ \hline\hline  
   1   & $ \rule{0cm}{4mm}       x_1 - x_3$        & 20/9 \\ \hline 
   2   & $ \rule{0cm}{4mm} -2 + 3(x_1+x_3)$        & \phantom{2}8/3 \\ \hline  
   3   & $ \rule{0cm}{4mm} 2 - 7(x_1+x_3) 
                      + 8(x_1^2+x_3^2)+4 x_1 x_3$ & 32/9 \\ \hline 
   4   & $ \rule{0cm}{4mm} (x_1 - x_3)(1 - 4/3 (x_1+x_3) 
                                       $ & 40/9 \\ \hline 
   5   & $ \rule{0cm}{4mm} 2-7(x_1+x_3) 
                      +14/3(x_1^2+x_3^2) +14x_1x_3$& 14/3 \\ \hline 
  \end{tabular}
 \end{center}
 \caption[]{Eigenfunctions and reduced anomalous dimensions for
       helicity 1/2 baryons. 
 \label{tab:spin1/2}}
\end{table}
%%%%%%%%%%%%%%%%%%%%%%%%%%%%%%%%%%%%%%%%%%%%%%%%%%%%%%%%%%%%%%%%%%%%%%%%

In \cite{BoK96} the nucleon \da\ was found to have the simple form 
\begin{equation}
  \hspace{-1cm}
  \phi^N_{123} (x,\mu_0) = \phi_{\rm AS}(x) \left[ 1 + \frac{3}{4} 
    \tilde \phi_{123}^{1}(x) + \frac{1}{4} \tilde \phi_{123}^{2}(x)
  \right] = 60x_1 x_2 x_3 \, \left[ 1 + 3 x_1 \right]. 
  \label{phiFIT}
\end{equation}
The nucleon wave function is fully specified now and, before turning
to the discussion of the hyperon \das, we note in passing that the
probability of the nucleon's valence Fock state is 0.17.

A suitable hyperon \da\ is constructed by taking (\ref{phiFIT}) and, 
in order to incorporate the empirically known breaking of $\su3$ 
symmetry, multiplying it with a factor
\begin{equation}
  \exp\left(-\frac{a_{(8)}^2 m_s^2}{x_j}\right)
  \label{BHLmassexp}
\end{equation}
whenever the quark $j$ is a strange one. That factor bears resemblance to
the BHL exponential \cite{BHL83}. $a_{(8)}$ is  the transverse size 
parameter already introduced in (\ref{BLHMOmega}) and $m_s$ is a still
to be adjusted parameter related to the strange quark mass. 
Explicitly our hyperon \das\ read
\EquSubNumb
\begin{align}
  \label{phiSig}
  \phi^{\Sigma}_{123}(x,\mu_0) & =
      N_{\Sigma}\,\phi^N_{123}(x)\,
      \exp\left(-\frac{a_{(8)}^2 m_s^2}{x_3}\right) \\
  \label{phiLam}
  \phi^{\Lambda}_{123}(x,\mu_0) & =
     \frac{1}{3} N_{\Lambda}\,\left( \phi^N_{123}(x) + 
            2\,\phi^N_{321}(x)\right)\,
      \exp\left(-\frac{a_{(8)}^2 m_s^2}{x_3}\right) \\
  \label{phiXi}
  \phi^{\Xi}_{123}(x,\mu_0) & =
      N_{\Xi}\,\phi^N_{123}(x)\,\exp\left(-a_{(8)}^2 m_s^2\left[\frac{1}{x_1}
      +\frac{1}{x_2}\right]\right)\,. 
\end{align}
\NoEquSubNumb
\hspace*{-0.2cm}The constants $N_{B_8}$ ensure the correct normalizations (see
(\ref{norm})) of the hyperon \das\ ($N_{B_8}=1$ for $m_s=0$).
The particular combination of $\phi_N$'s appearing in the $\Lambda$ case
(\ref{phiLam}) is required by $\su3$ symmetry. 

In order to take into account evolution properly we expand the \das\
(\ref{phiSig})-(\ref{phiXi}) upon the eigenfunctions of
the evolution kernel (see (\ref{DAentw})) up to terms of order $n=5$. In
Tab.\ \ref{tab:Octet-DAs} we quote four sets of expansion
coefficients $B_n$ corresponding to the following scenarios: Set 1 is
obtained from $m_s = 0$, set 2 from $m_s = 150$ MeV (current strange
quark mass) and set 4 from $m_s = 480$ MeV (constituent strange quark
mass). The intermediate set 3 corresponds to $m_s = 350\,\mathrm{MeV}
\approx (480^2 - 330^2)^{1/2}$ MeV (difference between the squares of 
strange and light constituent quark masses). This value appears if
(\ref{BHLmassexp}) is interpreted as the ratio of the BHL exponentials
for a hyperon and the nucleon \da. 

%
%%%%%%%%%%%%%%%%%%%%%%%%%%%%%%%%%%%%%%%%%%%%%%%%%%%%%%%%%%%%%%%%%%%%%%%
%    T A B L E   2  :  E X P A N S I O N   C O E F F I C I E N T S    %
%%%%%%%%%%%%%%%%%%%%%%%%%%%%%%%%%%%%%%%%%%%%%%%%%%%%%%%%%%%%%%%%%%%%%%%
%
\begin{table}
  \begin{tabular}{|l||r|r|r|r|r||r|r|r|r|r|} \hline
    & \multicolumn{5}{|c||}{Set 1 ($m_s = 0$ )} 
    & \multicolumn{5}{|c|} {Set 3 ($m_s = 350$ MeV)} \\ \hline 
    & $B_1\;\;$ & $B_2\;\;$ & $B_3\;\;$ & $B_4\;\;$ & $B_5\;\;$ 
    & $B_1\;\;$ & $B_2\;\;$ & $B_3\;\;$ & $B_4\;\;$ & $B_5\;\;$ \\ \hline\hline
    $\Sigma$  &  0.750 & 0.250 &  0.000 &  0.000 & 0.000 
              &  0.216 & 0.394 & -0.293 & -0.914 & 0.241 \\ \hline
    $\Lambda$ & -0.250 & 0.250 &  0.000 &  0.000 & 0.000 
              & -0.721 & 0.389 & -0.150 & -0.574 & 0.093 \\ \hline
    $\Xi$     &  0.750 & 0.250 &  0.000 &  0.000 & 0.000 
              &  1.106 & 0.050 & -0.282 &  1.717 &-0.498 \\ \hline\hline        
    & \multicolumn{5}{|c||}{Set 2 ($m_s = 150$ MeV)} 
    & \multicolumn{5}{|c|} {Set 4 ($m_s = 480$ MeV)} \\ \hline 
    & $B_1\;\;$ & $B_2\;\;$ & $B_3\;\;$ & $B_4\;\;$ & $B_5\;\;$ 
    & $B_1\;\;$ & $B_2\;\;$ & $B_3\;\;$ & $B_4\;\;$ & $B_5\;\;$ \\ \hline\hline
    $\Sigma$  &  0.623 & 0.284 & -0.085 & -0.285 & 0.065 
              & -0.118 & 0.484 & -0.404 & -1.173 & 0.358 \\ \hline
    $\Lambda$ & -0.360 & 0.282 & -0.048 & -0.195 & 0.027 
              & -1.022 & 0.478 & -0.182 & -0.650 & 0.127 \\ \hline
    $\Xi$     &  0.831 & 0.201 & -0.083 &  0.389 &-0.129
              &  1.338 &-0.068 & -0.384 &  2.943 &-0.775 \\ \hline        
  \end{tabular}
  \caption[]{Expansion coefficients $B_n(\mu_0)$ of octet baryon \das\ for
    various values of the parameter $m_s$.
\label{tab:Octet-DAs}}
\end{table}

\begin{figure}
\setlength{\unitlength}{1mm}
\begin{picture}(160,90)
 \put( 20,  45){\psfig{figure=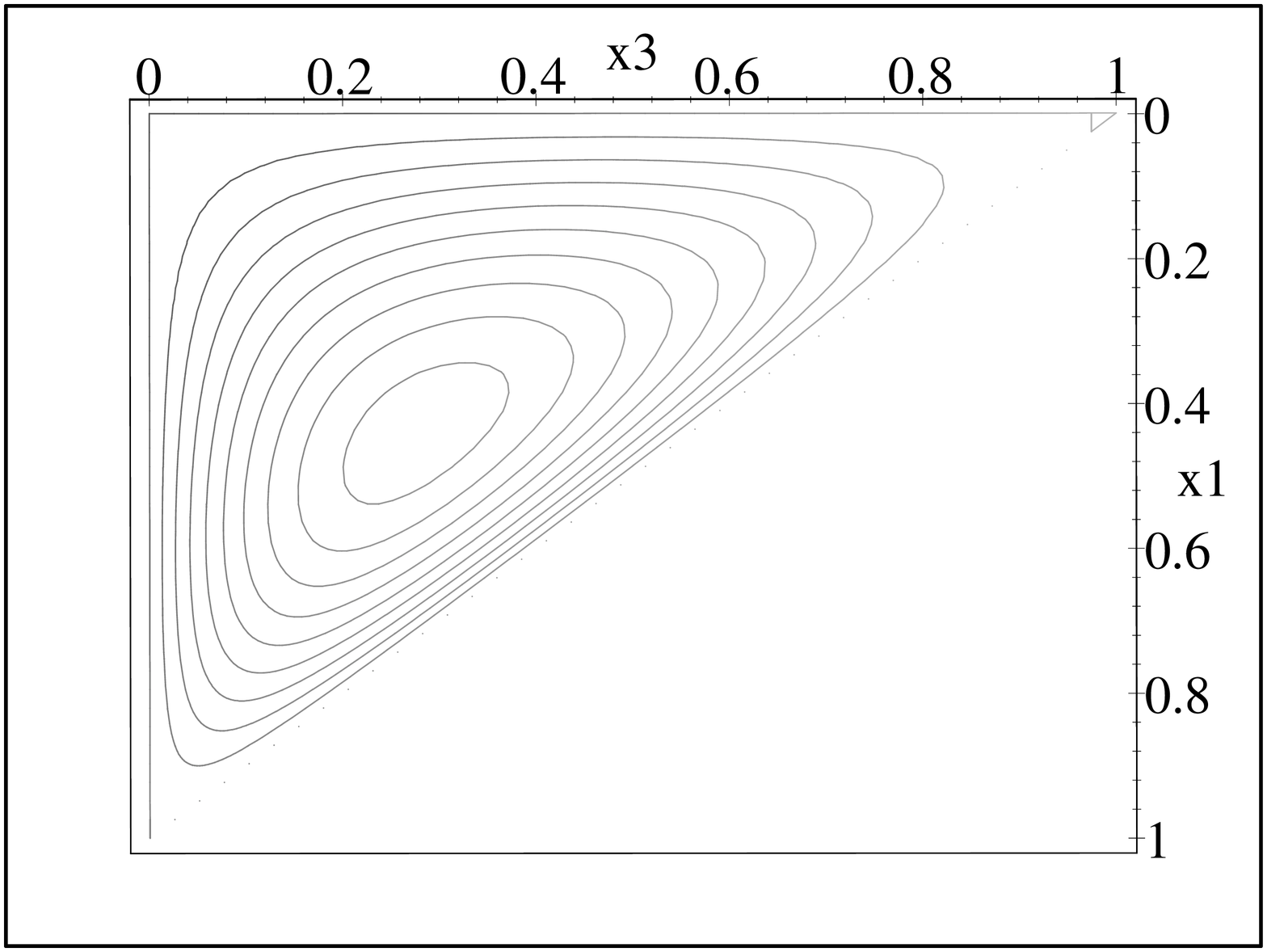,%
       bbllx=35pt,bblly=35pt,bburx=575pt,bbury=755pt,%
       height=4.5cm,angle=-90} }
 \put( 60,  60){\Large $N$}
 \put( 80,  45){\psfig{figure=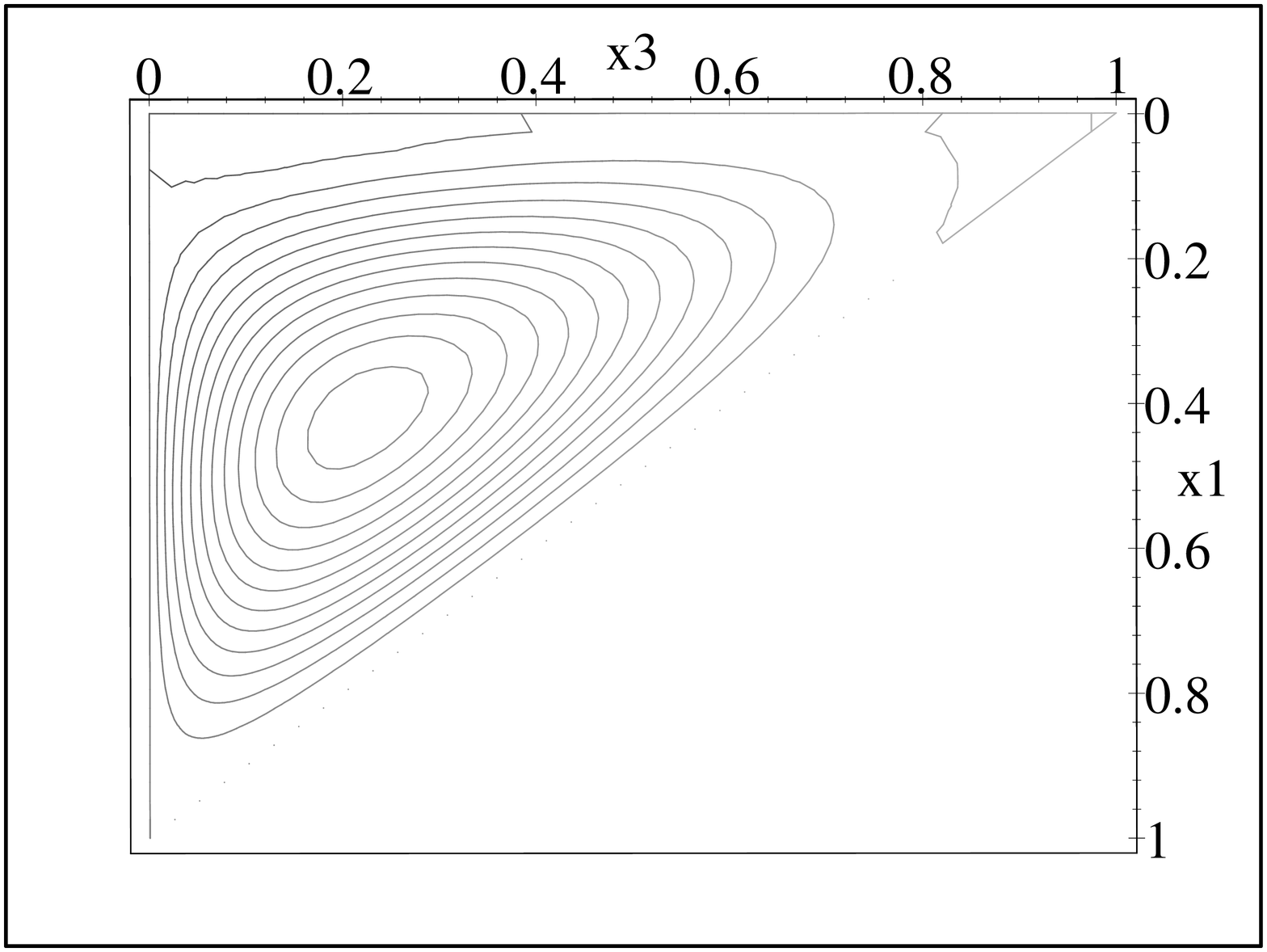,%
       bbllx=35pt,bblly=35pt,bburx=575pt,bbury=755pt,%
       height=4.5cm,angle=-90} }
 \put(120,  60){\Large $\Xi$}
 \put( 20,   0){\psfig{figure=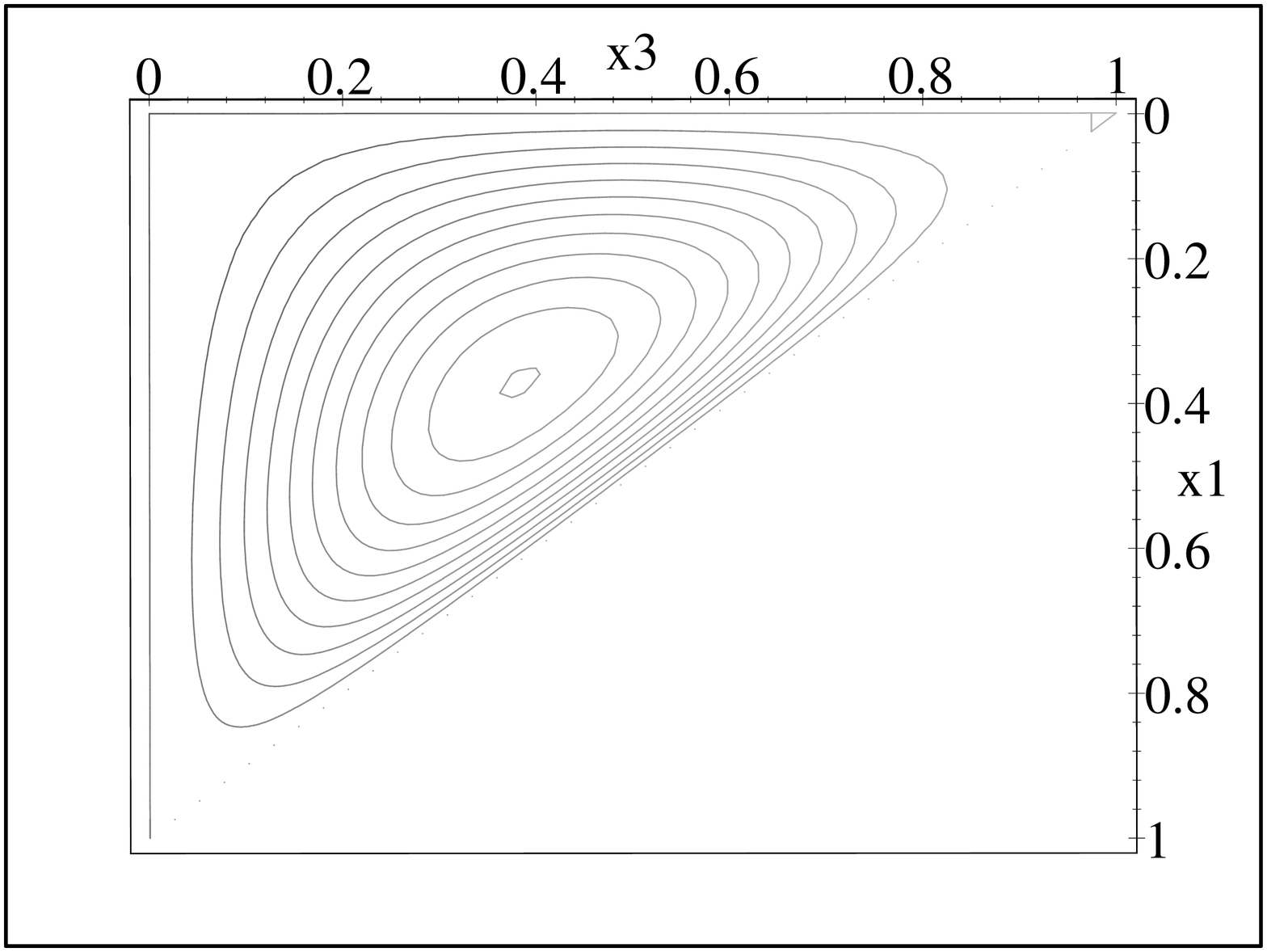,%
       bbllx=35pt,bblly=35pt,bburx=575pt,bbury=755pt,%
       height=4.5cm,angle=-90} }
 \put( 60,  15){\Large $\Sigma$}
 \put( 80,   0){\psfig{figure=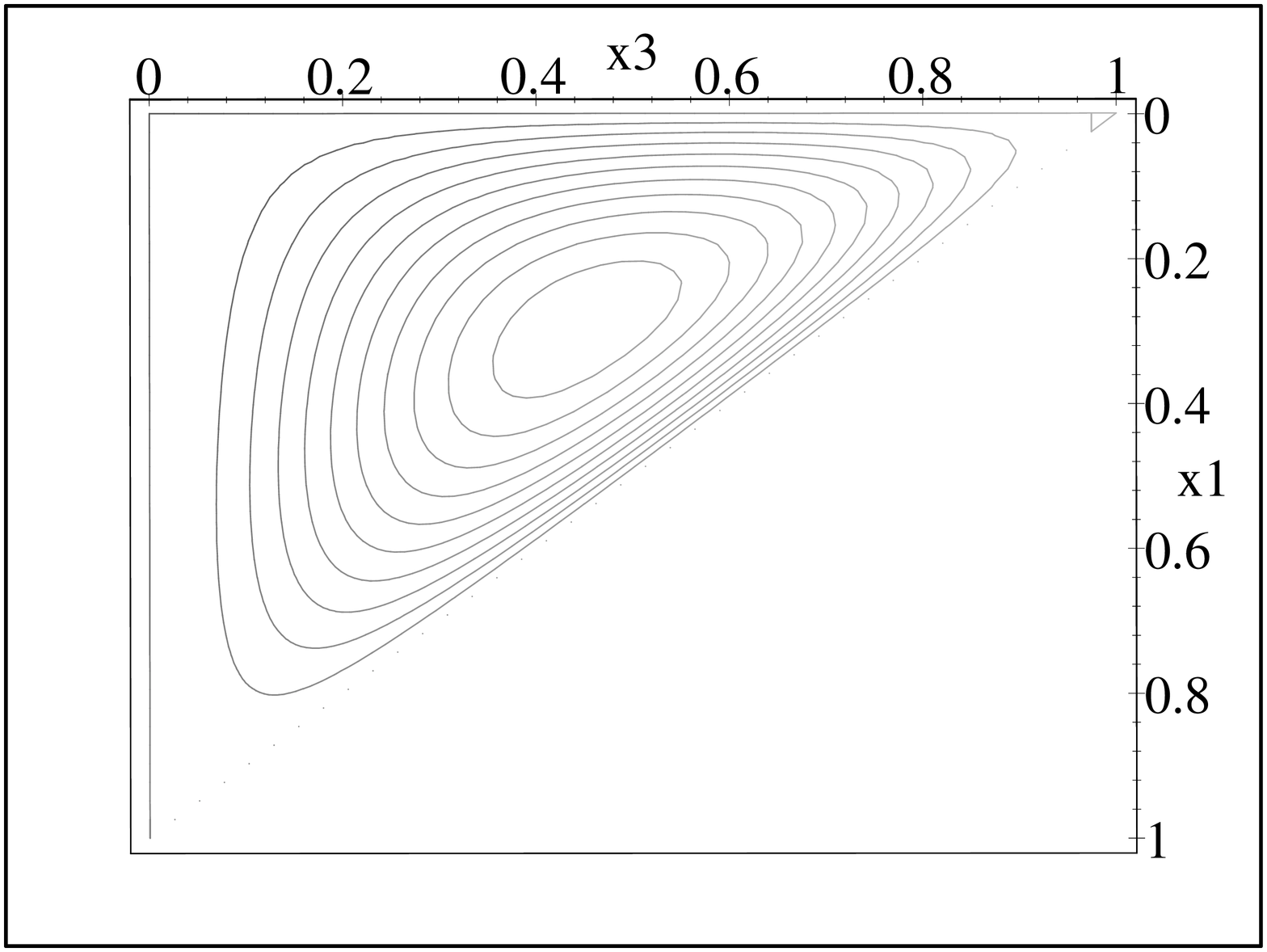,%
       bbllx=35pt,bblly=35pt,bburx=575pt,bbury=755pt,%
       height=4.5cm,angle=-90} }
 \put(120,  15){\Large $\Lambda$}
\end{picture}
 \caption[]{Contour plots of octet baryon \das\ for $m_s = 480$ MeV
 and at a scale of $1$ GeV.
  \label{fig:DAs} }
\end{figure}

In Fig.\ \ref{fig:DAs} we show contour plots of the four octet baryon 
\das\ for $m_s = 480$ MeV in order to illustrate the effect of the 
mass exponential (\ref{BHLmassexp}). It can be seen that,
compared to the nucleon case, the maxima of the $\Sigma$ and $\Lambda$
\das\ are shifted to the right, i.e.\ to larger $x_3$ values, whereas that of
the $\Xi$ \da\ is shifted to the left. On the average, $s$ quarks carry 
larger momentum fractions than $d$ quarks if $m_s > 0$.
%%%%%%%%%%%%%%%%%%%%%%%%%%%%%%%%%%%%%%%%%%%%%%%%%%%%%%%%%%%%%%%%%%%%%%%
%    F I G U R E   1  :  D I S T R I B U T I O N  A M P L I T U D E S %
%%%%%%%%%%%%%%%%%%%%%%%%%%%%%%%%%%%%%%%%%%%%%%%%%%%%%%%%%%%%%%%%%%%%%%%
%
%

Distribution amplitudes for the octet baryons are also discussed in
\cite{COZ89a}. As compared to our ones these QCD sum rule based \das\ 
are strongly concentrated in the end-point regions and exhibit a more 
significant breaking of $\su3$ symmetry. Moreover, they reveal three 
pronounced maxima near the end-points. In summary, the \das\ proposed in
\cite{COZ89a} are very different from ours and also from the asymptotic
\da\ at experimentally accessible scales. As our \das\, but to a much
greater extend, they possess the property that, on the average, a $u$
($s$) quark in the proton or the $\Sigma^+$ ($\Xi^-$, $\Lambda$) with 
the same helicity as its parent baryon carries the largest
fraction of the baryon momentum.

%
%%%%%%%%%%%%%%%%%%%%%%%%%%%%%%%%%%%%%%%%%%%%%%%%%%%%%%%%%%%%%%%%%%%%%%%
\section{The wave functions of the decuplet baryons \label{Sect:dec}}
%%%%%%%%%%%%%%%%%%%%%%%%%%%%%%%%%%%%%%%%%%%%%%%%%%%%%%%%%%%%%%%%%%%%%%%
%
\setcounter{equation}{0}
In analogy to (2.1) we write the valence Fock states of the
decuplet baryons $\Delta$ and $\Sigma^{\star}$ with helicity
$\lambda_{B_{10}}=1/2$ as%
\footnote{We omit the discussion of the decuplet states $\Xi(1530)$
and $\Omega^-$ since the perturbative approach is not applicable to
the decays $\jp \to \Xi(1530)\overline \Xi(1530)$ due to almost zero
momentum transfer. The decay $\jp \to \Omega^-\overline{\Omega}{}^+$ 
is even kinematically forbidden. The decay into $\Sigma^*$ is the
borderline case for the application of the perturbative approach.}   
\EquSubNumb
\begin{alignat}{2}
%
%--------- Delta++ -------------
%
  \label{Del++state}
  |\Delta^{++},+ \,\rangle\, & 
=
  \frac{\varepsilon _{a_{1}a_{2}a_{3}}}{\sqrt{2}} 
  \int
  [{\rm d}x]
  [{\rm d}^{2}{\bf k_{\perp}}] & \,
    \Psi^{\Delta} _{123}\:|\,u_+^{a_1}\; u_-^{a_2}\; u_+^{a_3}\,\rangle 
   \phantom{  
   + \Psi^{\Delta} _{213}\:|\,u_-^{a_1}\; u_+^{a_2}\; d_+^{a_3}\,\rangle
            \Bigr\} \;\;\;}\\
%
%--------- Delta+ -------------
%
  \label{Del+state}
  |\;B_{10}\;,+ \,\rangle\, & 
=
  \frac{\varepsilon _{a_{1}a_{2}a_{3}}}{\sqrt{3!}} 
  \int
  [{\rm d}x]
  [{\rm d}^{2}{\bf k_{\perp}}] &
  \Bigl\{ 
         \Psi^{B_{10}} _{123}\:|\,f_{1+}^{a_1} f_{1-}^{a_2} f_{2+}^{a_3}\,\rangle
       + \Psi^{B_{10}} _{213}\:|\,f_{1-}^{a_1} f_{1+}^{a_2} f_{2+}^{a_3}\,\rangle
  \phantom{ \Bigr\} \;\;\; }
\nonumber \\
       & & +
         \Psi^{B_{10}} _{132}\:|\,f_{1+}^{a_1} f_{1+}^{a_2} f_{2-}^{a_3}\,\rangle
  \Bigr\} \; .\;\;  
\end{alignat}
\NoEquSubNumb
\hspace{-0.3cm}For the $\Delta^+$ ($\Sigma^{*-}$) $f_1$ represents an
$u$ ($d$) quark and $f_2$ a $d$ ($s$) one. Again, the states of the 
$\Delta^0$, $\Delta^-$ and $\Sigma^{\star +}$ are obtained from those 
of the $\Delta^+$, $\Delta^{++}$ and $\Sigma^{\star -}$ by exchanging 
$u \leftrightarrow d$, respectively. The generalization to the other 
helicity states is trivial. There is obviously only one independent 
scalar wave function in the case of the $\Delta$ and we assume that 
the same shall apply to the $\Sigma^*$. The scalar wave functions 
$\Psi^{B_{10}}_{123}$ of the decuplet baryons 
$B_{10}\,(=\Delta,\Sigma^{\star})$ are parameterized in a 
fashion similar to the octet baryon case: 
\begin{equation}
  \Psi^{B_{10}} _{123}(x,{\bf k_{\perp}})
=
  \frac{f_{(10)}(\mu_F)}{24\sqrt{2}}\,
  \phi^{B_{10}}_{123}(x,\mu _{F})\,
  \Omega_{(10)} (x,{\bf k_{\perp}})\:. 
\label{PsiDansatz}
\end{equation}
$f_{(10)}$ is assumed to be equal for all members of the baryon decuplet.   
We again adopt the form (\ref{BLHMOmega}) for the transverse momentum 
dependent part of the wave function, $\Omega_{(10)}$, with the transverse 
size parameter $a_{(8)}$ replaced by $a_{(10)}$. 

Decuplet baryons in helicity 1/2 states have the same eigenfunctions
of the evolution kernel and the same anomalous dimensions as the octet
baryons (see Tab.\ \ref{tab:spin1/2}). Evolution of the helicity 3/2
baryons is, on the other hand, different \cite{BrL80}. We refrain from
giving details on that case here since we do not consider such baryons. 

In the particular case of the $\Delta$ it is tempting to use a
completely permutation symmetric \da\ since the $\Delta$ is composed of
are three light quarks in symmetric spin and 
flavour states. Noting that the permutation symmetric part of the 
nucleon \da\ (\ref{phiFIT}) is just equal to the asymptotic \da, we 
take $\phi_{123}^{\Delta}(x) = \phi_{\mathrm{AS}}(x)$ and construct 
$\phi^{\Sigma^{\star}}(x)$ analogously to $\phi^{\Sigma}(x)$ 
\begin{equation}
  \label{phiSigst}
  \phi^{\Sigma^{\star}}_{123}(x) =
      N_{\Sigma^*}\,\phi_{\mathrm{AS}}(x)\,
      \exp\left(-\frac{a_{(10)}^2 m_s^2}{x_3}\right)\,. 
\end{equation}
The evolution of the $\Sigma^*$ \da\ is treated as in the case of 
the octet baryons by expanding (\ref{phiSigst}) upon the eigenfunctions 
$\tilde \phi^{n}_{123}(x)$ up to $n=5$. In distinction from the
octet baryon case, the QCD sum rule based \das\ for the
decuplet baryons \cite{ste94,far89} are not very  different from the 
ones we are proposing.

%
%%%%%%%%%%%%%%%%%%%%%%%%%%%%%%%%%%%%%%%%%%%%%%%%%%%%%%%%%%%%%%%%%%
\section{Decays of the $\jp$ into octet baryons}
%%%%%%%%%%%%%%%%%%%%%%%%%%%%%%%%%%%%%%%%%%%%%%%%%%%%%%%%%%%%%%%%%%
%
\setcounter{equation}{0}
Now, with the model wave functions at hand, we can calculate the decay 
width of $\jp$ into octet baryon-antibaryon pairs within the modified
perturbative approach. The helicity amplitudes of these processes may be 
decomposed covariantly to read
\begin{equation}
  {\mathcal M}^{B_8}_{\lambda_1 \lambda_2 \lambda} = 
       \overline u_{B_8}(p_1,\lambda_1)
  \left[ {\mathcal B}^{B_8} \, \gamma_{\mu} + {\mathcal C}^{B_8}\,
    \frac{(p_1 - p_2)_{\mu}}{2\,m_{B_8}} \right] v_{B_8}(p_2,\lambda_2) \, 
  \epsilon^{\mu}(\lambda)    
  \label{covzerl}
\end{equation}
where $m_{B_8}$ is the mass of an octet baryon. $p_1$ ($p_2$) and 
$\lambda_1$ ($\lambda_2$) denote the momentum and the helicity of an 
octet baryon (antibaryon), respectively. $u_{B_8}$ and $v_{B_8}$
are their spinors (normalized as 
$\overline u_{B_8} u_{B_8} = 2m_{B_8}$) and $\epsilon$ is the
polarization vector of the $\jp$. In a leading twist perturbative 
approach the helicity amplitudes are determined by the invariant 
${\mathcal B}^{B_8}$ (${\mathcal C}^{B_8}=0$). Implicitly this ensures
hadronic helicity conservation. The $\jp \to B_8\overline B_8$ decay 
width reads 
\begin{equation}
   \label{GamB}
   \Gamma(\jp \to O \overline O)  =
   \frac{\rho_{\mathrm{p.s.}}(m_{B_8}/\mpsi)}{48\,\pi\,\mpsi}
      \:\sum |\,{\mathcal M}^{B_8}_{\lambda_1 \lambda_2 \lambda}|^2\,
\end{equation}
in the $\jp$ rest frame ($\mpsi$ is the $\jp$ mass). If the $\jp$ is 
produced in $e^+e^-$ annihilations it is transversely polarized with 
respect to the beam direction and the angular distribution of the 
baryons emitted in the $\jp$ rest frame then exhibits a
$1+\cos^2{\theta}$ dependence (up to corrections of 
${\mathcal O}(m_{B_8}^2/\mpsi^2))$ characteristic of perturbative QCD 
\cite{BrL81}. The function $\rho_{\mathrm{p.s.}}$ in (\ref{GamB}) is 
the usual phase space factor  
\begin{equation}
   \rho_{\mathrm{p.s.}}(z)  = \sqrt{1-4 z^2}\, .
   \label{fphsp}
\end{equation}

As we said in the introduction we are going to calculate the invariant
${\mathcal B}^{B_8}$ within the modified perturbative approach proposed in
\cite{BoSLi}, thus generalizing our analysis of the $\jp$ decay into
nucleon-antinucleon pairs \cite{BoK96}. As in previous perturbative
calculations \cite{BrL81,COZ89b,BeS92} the $\jp$ meson is
treated as a non-relativistic $c\overline c$ system and 
${\mathcal O}\,(v^2/c^2)$ corrections are neglected. In contrast to
the decays of $P$-wave charmonium \cite{BKS96} the $c\overline{c}g$ 
Fock state is suppressed in exclusive decays by inverse powers of 
the large scale which is provided by the $c$-quark mass $m_c$ 
\cite{bar81,bod95}, relative to the $c\overline c$ Fock state and is, 
therefore, neglected in our analysis. The use of $m_c$, strictly
speaking $2 m_c$, as the large scale rather than the charmonium mass 
is consistent with the neglect of relativistic corrections. It is 
also well in the spirit of a perturbative approach since in the 
internal $c$-quark propagators the $c$-quark mass appears. 
The $\jp$ state is written in a covariant fashion
\begin{equation}
  |\,\jp; \,q,\lambda\,\rangle \,=\, \frac{\delta_{ab}}{\sqrt{3}}\,
      \left (\frac{f_{\psi}}{2\,\sqrt{6}} \right )\,
        \frac{1}{\sqrt{2}} (\qsla+\mpsi) \esla(\lambda)
  \label{Jpsistate}
\end{equation}
where $a$ and $b$ are colour indices and $f_{\psi}$ is the $\jp$ decay 
constant being related to the $\jp$ wave function at the origin of the
configuration space. Merely that part of the wave function is, to a
reasonable approximation, required in the calculation of ${\mathcal B}$ 
since, as we already mentioned in the introduction, the $c$ and the 
$\overline c$ quark only annihilate if their mutual distance is less 
than about $1/m_c$ \cite{app75} which is smaller than the $\jp$ radius
\cite{buc81}. $\mpsi$ is replaced by $2m_c$ in the calculation of 
${\mathcal B}_{3g}^{B_8}$ and the baryon masses are neglected. Only
the phase space factor in (\ref{GamB}) is evaluated with the physical 
masses, $\mpsi$ and $m_{B_8}$. 

\begin{figure}
\[
 \psfig{figure=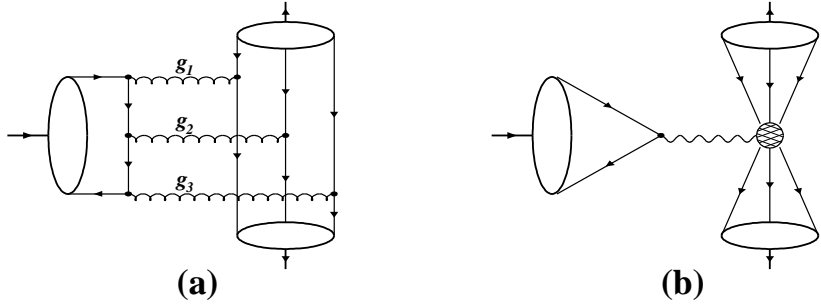,%
       bbllx=50pt,bblly=40pt,bburx=800pt,bbury=285pt,%
       height=4cm,clip=} \]
 \caption[dummy0]{Feynman graphs for the $\jp$ decay ointo 
       baryon-antibaryon. a) Three-gluon contribution (graphs with
       permutated gluon lines are not shown), b) electromagnetic contribution.
  \label{fig:graphs} }
\end{figure}

The invariant ${\mathcal B}^{B_8}$ receives its dominant
contribution from the graphs with three intermediate gluons,
see Fig.~\ref{fig:graphs}. Within the modified perturbative
approach the three-gluon contribution ${\mathcal B}_{3g}^{B_8}$ to the
$\jp$ decay into $B_8\overline B_8$ is of the form 
\bea
\label{B3gOct} 
  {\mathcal B}^{B_8}_{3g} &=& \frac{f_{\psi}}{2\sqrt{6}}\,
     \int [{\rm d}x] [{\rm d}x'] \int\,\frac{{\rm d}^2{\bf b}_1}{(4\pi)^2}\,
     \frac{{\rm d}^2{\bf b}_3}{(4\pi)^2} \,
     \hat T_H(x,x',{\bf b})\,\exp[-S(x,x',{\bf b},2m_c)]\\ 
   &\times& 
  \big[ \hat\Psi^{B_8}_{123}(x,{\bf b}) \hat\Psi^{B_8}_{123}(x',{\bf b})  
         + \frac{1}{2}
      \big(\hat\Psi^{B_8}_{123}(x ,{\bf b} ) +  \hat\Psi^{B_8}_{321}(x ,{\bf b} )
                                                               \big)\! 
      \big(\hat\Psi^{B_8}_{123}(x',{\bf b})  
         + \hat\Psi^{B_8}_{321}(x',{\bf b}) \big) 
     \big] \nn
\eea   
for $B_8 = N,\Sigma,\Xi$. This contribution is the same for all octet baryons 
belonging to the same isospin multiplet. Since the form of the
$\Lambda$ Fock state (\ref{Lamstate}) is slightly different the three-gluon
contribution reads 
\bea
\label{B3gLam} 
  {\mathcal B}^{\Lambda}_{3g} &=& \sqrt{\frac{3}{2}}\frac{f_{\psi}}{2}\,
     \int [{\rm d}x] [{\rm d}x'] \int\,\frac{{\rm d}^2{\bf b}_1}{(4\pi)^2}\,
     \frac{{\rm d}^2{\bf b}_3}{(4\pi)^2} \,
     \hat T_H(x,x',{\bf b})\,\exp[-S(x,x',{\bf b},2m_c)]\\ 
  &\times& \,
  \big[ \hat\Psi^{\Lambda}_{123}(x,{\bf b}) \hat\Psi^{\Lambda}_{123}(x',{\bf b})  
         + \frac{1}{2}
      \big(\hat\Psi^{\Lambda}_{123}(x ,{\bf b} ) 
         -  \hat\Psi^{\Lambda}_{321}(x ,{\bf b} )  \big)\! 
      \big(\hat\Psi^{\Lambda}_{123}(x',{\bf b}) 
         -  \hat\Psi^{\Lambda}_{321}(x',{\bf b})   \big)
     \big] 
     \, \nonumber 
\eea
%
%%%%%%%%%%%%%%%%%%%%%%%%%%%%%%%%%%%%%%%%%%%%%%%%%%%%%%%%%%%%%%%%%%%%%%%
%    F I G U R E   2  :  F E Y N M A N   G R A P H S                  %
%%%%%%%%%%%%%%%%%%%%%%%%%%%%%%%%%%%%%%%%%%%%%%%%%%%%%%%%%%%%%%%%%%%%%%%
%
%
\hspace*{-0.3cm} in the case of the $\Lambda$. The representation of 
${\mathcal B}_{3g}$ as a convolution of wave functions and a hard 
scattering amplitude $\hat{T}_H$ can formally be derived by using the 
methods described in detail by Botts and Sterman \cite{BoSLi}. The 
${\bf b}_i$, canonically conjugated to the transverse momenta 
${\bf k_{\perp}}_i$, are the quark separations in the transverse 
configuration space. ${\bf b}_1$ and ${\bf b}_3$ correspond to the 
locations of quarks 1 and 3 in the transverse plane relative to quark
2 and ${\bf b}_2 = {\bf b}_1 - {\bf b}_3$. $\hat\Psi_{ijk}^{B_8}$ 
represents the Fourier transform of the wave function 
$\Psi_{ijk}^{B_8}$ (see Sect.\ 2).  

$\hat{T}_H$ is the Fourier transform of the usual momentum space hard
scattering amplitude to be calculated from the Feynman graphs shown in
Fig.\ \ref{fig:graphs}a. Up to corrections of order $\al^4$, 
$m^2_{B_8}/(4 m_c^2)$ and $b^2/(4m_c^2)$ the hard scattering amplitude 
in ${\bf b}$ space reads  
\begin{eqnarray}
  \hat T_H(x,x',{\bf b}) 
    & = &
     -\frac{5}{27}2^{12} \,f_{\psi}\, m_c^6\,
     \frac{(x_1 x'_3 + x_3 x'_1)}
     {[\tilde q_1^2 + \tilde g_1^2]
      [\tilde q_3^2 + \tilde g_3^2]} \,
     (\prod_{i=1}^3 \alpha_S(t_i))\,
     \int {\rm d}^2{\bf b}_0\,\hspace{1cm}
     \nonumber \\
    & \times &
     \left[ \frac{i\pi}{2} \Ha{0}(\tilde g_1\,|{\bf b}_1 + {\bf b}_0|\,)
     - \BK{0}(\tilde q_1\,|{\bf b}_1 + {\bf b}_0|\,) \right] \,
     \frac{i\pi}{2} \Ha{0}(\tilde g_2 b_0) \,
     \nonumber \\
    & \times &
     \left[ \frac{i\pi}{2} \Ha{0}(\tilde g_3\,|{\bf b}_3 + {\bf b}_0|\,)
     - \BK{0}(\tilde q_3\,|{\bf b}_3 + {\bf b}_0|\,) \right] \: .   
  \label{THfour2}
\end{eqnarray}
The quantities
\begin{equation}
  \tilde q_i^2 = 2[ x_i\,(1-x_i') + (1-x_i)\,x_i']\,m_c^2
      \, , \hspace{1cm}
  \tilde g_i^2 = 4 x_i x_i'\,m_c^2 \, \hspace{0.5cm}
  \label{longscaljpsi}
\end{equation}
represent the virtualities of the internal quarks and gluons at zero 
transverse momenta. Since, as shown in \cite{BoK96}, the hard 
scattering amplitude only depends on the sum of transverse momenta, 
${\bf k}_{\perp i}+{\bf k'}_{\perp i}$, the transverse separation of 
any two quarks inside the baryon is the same as that of the 
corresponding antiquarks inside the antibaryon  (${\bf b}_i = {\bf b'}_i$). 
Physically, this property of the hard scattering amplitude means that 
baryon and antibaryon are created with identical transverse
configurations of the quarks and antiquarks, respectively. The 
auxiliary variable ${\bf b}_0$ in (\ref{THfour2}) serves as a Lagrange
multiplier to the constraint $\sum {\bf k}_i+{\bf k'}_i = {\bf 0}$. 
Since the virtualities of the gluons are timelike, $\hat T_H$ includes 
complex-valued Hankel functions $\Ha{0}$ that are related to the 
usual modified Bessel functions $\BK{0}$, appearing for space-like 
propagators, by analytic continuation. 

The Sudakov factor $\exp[-S]$ entering (\ref{B3gOct}) and 
(\ref{B3gLam}) takes into account those gluonic radiative corrections 
not accounted for in the QCD evolution of the wave function as well 
as the renormalization group transformation from the factorization 
scale $\mu_F$ to the renormalization scales $t_i$ at which the hard 
amplitude $\hat T_H$ is evaluated. The Sudakov factor, originally 
derived by Botts and Sterman \cite{BoSLi} and later on slightly 
improved, can be found for instance in \cite{DJK95}. The renormalization 
scales $t_i$ are defined in analogy to the case of electromagnetic 
form factors \cite{BoSLi} as the maximum scale of either the 
longitudinal momentum or the inverse transverse separation associated 
with each of the gluons
\begin{equation}
   t_1 = \max(\tilde q_1, \tilde g_1, 1/b_3) \,, \hspace{0.5cm}
   t_2 = \max(\tilde g_2, 1/b_2)             \,, \hspace{0.5cm}
   t_3 = \max(\tilde q_3, \tilde g_3, 1/b_1) \,.
   \label{tijpsi}
\end{equation}

Infrared cut-off parameters $\tilde b_i$ appear in the Sudakov factor
which are naturally related to, but not uniquely determined by the
mutual separations of the three quarks \cite{CoS81}. Following
\cite{wubo} we chose $\tilde b_i = \tilde b = \max\{b_1,b_2,b_3\}$. 
With this ``MAX'' prescription the three-gluon contribution 
${\cal B}_{3g}^{B_8}$ is unencumbered by $\alpha_S$ singularities in 
the soft end-point regions. As a consequence of the regularizing 
power of the ``MAX'' prescription, the perturbative contribution 
saturates in the sense that the results become insensitive to the 
inclusion of the soft regions. A saturation as strong as possible 
is a prerequisite for the self-consistency of the perturbative 
approach. The infrared cut-off $\tilde b$ marks the interface 
betweeen the non-perturbative soft gluons, which are implicitly 
accounted for in the baryon wave function, and the contributions 
from soft gluons, incorporated in a perturbative way in the Sudakov 
factor. Obviously, the gliding factorization scale to be used in the 
evolution of the wave function, has to be chosen as $\mu_F=1/\tilde b$. 

As an inspection of (\ref{B3gOct}), (\ref{B3gLam}) and 
(\ref{THfour2}) reveals, a nine dimensional numerical integration has 
to be performed.%
\footnote{Taking into account relativistic corrections to the $\jp$
  wave function, i.e.\ its transverse momentum dependence, one would
  have to perform a 14 dimensional numerical integration which seems
  impossible with present day computers to a sufficient degree of
  accuracy.} 
Although this is a rather involved technical task it can be carried
through with sufficient accuracy if some care is put into it. The
numerical results are obtained from the wave functions discussed in
Sect.\ 2 and for the following values of the $\jp$ decay constant and
the $c$-quark mass: $f_{\psi}= 409\,\mev$, $m_c= 1.5\,\gev$. We 
evaluate $\al$ in the one-loop approximation with $n_f=4$ and 
$\Lambda_{QCD} =210\,\mev$ \cite{pdg96}.

%
%%%%%%%%%%%%%%%%%%%%%%%%%%%%%%%%%%%%%%%%%%%%%%%%%%%%%%%%%%%%%%%%%%%%%%%
%    F I G U R E   3  :  A L P H A _ S ^ {C R I T}                %
%%%%%%%%%%%%%%%%%%%%%%%%%%%%%%%%%%%%%%%%%%%%%%%%%%%%%%%%%%%%%%%%%%%%%%%
%
%
\begin{figure}
\[
 \psfig{figure=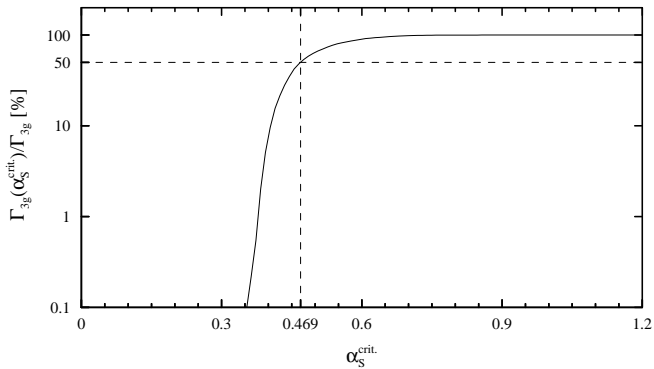,%
       bbllx=50pt,bblly=485pt,bburx=550pt,bbury=770pt,%
       height=5cm,clip=} \]
 \caption[]{Percental accumulation of the three-gluon contribution 
       to the width of the $N\overline N$ channel from
       regions of internal momenta where $\prod_{i=1}^3 \alpha_s(t_i)
       < (\alpha_s^\mathrm{crit}{})^3$.           
  \label{fig:alsc} }
\end{figure}  
Before turning to the detailed discussion of the numerical results for the
decay widths, we want to focus on an important feature of the 
modified perturbative approach. As mentioned in the introduction, 
one of the motivations for including the transverse hadronic structure 
and the Sudakov factor in the analysis is to achieve a theoretically 
self-consistent calculation in the sense that the bulk of the 
perturbative contribution is accumulated in regions where the strong 
coupling constant $\alpha_s$ is sufficiently small. A method to check 
whether or not this is the case is to set the integrand in 
(\ref{B3gOct}) or (\ref{B3gLam}) equal to zero in those regions where 
$\prod_{i=1}^3 \alpha_s(t_i) > (\alpha_s^\mathrm{crit}{})^3$ 
and to evaluate the three-gluon contribution to the decay width as 
a function of $\alpha_s^\mathrm{crit}$. In Fig.\ \ref{fig:alsc} we 
show the accumulation profile for the nucleon case; it is typical
of all baryons. Consulting Fig.\ \ref{fig:alsc}, one sees that almost
the entire result is accumulated in the comparatively narrow region 
of $\alpha_s$ between 0.4 and 0.6. The regions with 
$\prod_{i=1}^3 \alpha_s(t_i) < 0.469^3$ provide 50 \% of the total 
result. Hence, our calculation of the $\jp$ decay widths into 
octet baryon-antibaryon pairs is theoretically self-consistent. 

\begin{figure}
 \[ \psfig{figure=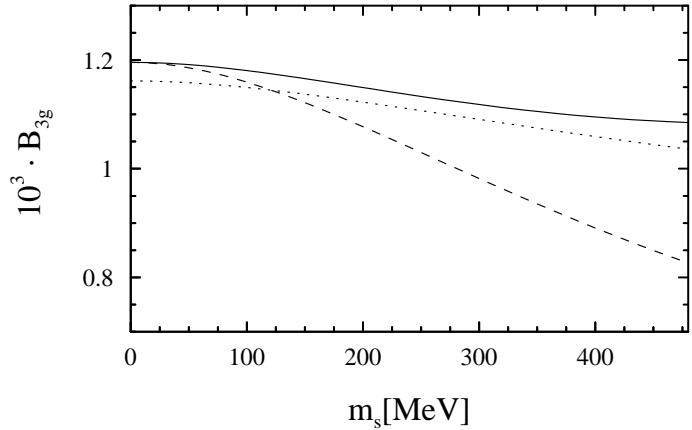,%
       bbllx=35pt,bblly=470pt,bburx=460pt,bbury=745pt,%
       height=6cm,clip=} \vspace*{-5mm} \]
 \caption[]{The three-gluon contribution ${\mathcal B}_{3g}^{B_8}$ 
   vs.\ the mass parameter $m_s$ for the decay channels 
   $\Sigma \overline\Sigma$ (solid), $\Xi \overline\Xi$
   (dashed) and $\Lambda \overline\Lambda$ (dotted). 
%($f_{\psi}=409\,$MeV, $m_c=\mpsi/2$, $\LQCD=220\,$ MeV).  
  \label{fig:B3g} }
\end{figure}

In Tab.\ \ref{tab:Gamma} we present our results for the $\jp \to
B_8\overline B_8$ decay widths using the \das\ discussed in Sect.\
2. For the sake of comparison we also expose the available experimental 
data \cite{pdg96}%
\footnote{
We use PDG averages throughout. The original data are from
\cite{eat84,DM2,hen87,Ant93a}.}. 
As can be seen from the results obtained with the \das\ termed set 1 
the phase space factor $\rho_{\mathrm {p.s.}}$  
%
%%%%%%%%%%%%%%%%%%%%%%%%%%%%%%%%%%%%%%%%%%%%%%%%%%%%%%%%%%%%%%%%%%%%%%%
%    T A B L E   3  :  R E S U L T S   F O R   W I D T H S            %
%%%%%%%%%%%%%%%%%%%%%%%%%%%%%%%%%%%%%%%%%%%%%%%%%%%%%%%%%%%%%%%%%%%%%%%
%
\begin{table}
 \begin{center}
  \begin{tabular}{|l||r|r|r|r|r|r|} \hline
      & \multicolumn{5}{|c|}{ $\Gamma_{3g}$ [eV]} & 
        Data \cite{pdg96} \\ \hline
    Channel & Set 1 & Set 2 & 
      Set 3 & Set 4 & $\phi_{\mathrm{AS}}\;$ & 
              $\Gamma_{\mathrm{exp}}$ [eV] \\ \hline\hline
    $p \overline p$                & 174 & 174 & 174 & 174 & 140 & 
                                        $186 \pm 14$   \\ \hline  
 $\Sigma^0 \overline {\Sigma}{}^0$ & 136 & 128 & 113 & 108 & 97.8&
                                        $110 \pm 15$   \\ \hline
    $\Lambda  \overline \Lambda$   & 140 & 133 & 117 & 107 & 99.7& 
                                        $117 \pm 14$   \\ \hline
    $\Xi^- \overline {\Xi}{}^+$    & 107 & 92.8& 62.5& 47.4& 46.9&
                                        $ 78 \pm 18$   \\ \hline        
  \end{tabular}
 \end{center}
 \caption[]{Results for the decay widths of $\jp$ into
    octet baryon-antibaryon pairs for the four sets of \das\ defined 
    by the expansion coefficients $B_n$ quoted in Tab.\ \ref{tab:Octet-DAs}
    ($f_{\psi}=409\,$ MeV, $m_c=1.5\,\gev$). For comparison we also 
    quote the experimental results and, in the column labelled 
    $\phi_{\mathrm{AS}}$, predictions evaluated with \das\
    constructed from the asymptotic \da\ instead from (\ref{phiFIT})
    (with $m_s=350\,$ MeV). 
\label{tab:Gamma}}
\end{table}
is an important but not sufficient element for the suppression of the hyperon   
channels. Since $m_s=0$ for these \das\ the differences in the
predictions for the widths, except for the $\Lambda \overline \Lambda$
case, are only due to $\rho_{\mathrm{p.s.}}$. For $m_s>0$ the
additional suppression of the end-point regions leads to smaller decay 
widths for the hyperon channels. In order to demonstrate the strength
of that reduction we display the invariant ${\mathcal B}_{3g}^{B_8}$ 
versus $m_s$ in Fig.\ \ref{fig:B3g}. The number of strange quarks
embodied in a given baryon is reflected in differently strong $m_s$ 
dependences of ${\mathcal B}_{3g}^{B_8}$. As inspection of 
Tab.\ \ref{tab:Gamma} brings to view, the phase space corrected 
three-gluon contributions nicely reproduce the experimental pattern of
the decay widths provided they are computed with the \das\ of set 3. 
The value of 350 \mev\ for the parameter $m_s$ used in the
construction of these \das\, appears reasonable, considering the 
interpretation of the mass factor (\ref{BHLmassexp}) as the BHL 
exponential \cite{BHL83}. Note that, as a consequence of the use of 
the $c$-quark mass in the calculation of ${\mathcal B}_{3g}^{B_8}$, 
the result for the decay width of the $p\overline p$ channel given in 
Tab.\ \ref{tab:Gamma} differs from that one reported in \cite{BoK96}.   
%
%%%%%%%%%%%%%%%%%%%%%%%%%%%%%%%%%%%%%%%%%%%%%%%%%%%%%%%%%%%%%%%%%%%%%%%
%    F I G U R E   4  :  B_{3g}  v e r s u s   m_s                    %
%%%%%%%%%%%%%%%%%%%%%%%%%%%%%%%%%%%%%%%%%%%%%%%%%%%%%%%%%%%%%%%%%%%%%%%
%

The \das\ defined by (\ref{phiFIT}-\ref{phiXi}) exhibit a little
asymmetry with respect to permutations of the $x_i$. In order to study
the importance of this asymmetry we also show in Tab.\ \ref{tab:Gamma} 
results computed with the asymptotic \da\ for the nucleon and with 
modifications of it for the hyperons that are constructed analogously 
to the \das\ described in Sect.\ 2. The asymmetry in the \das\ is seen
to increase the magnitudes of the decay widths by about 30 \% for the 
$\Xi\overline \Xi$ channel and about 20\% for the other channels while
the pattern of the predictions remains unchanged. 

Let us now assess the uncertainties of our predictions.
The value of the $\jp$ decay constant used by us is determined from the
leptonic $\jp$ decay width. Since the $\jp$ decay constant, or more
generally, the decay constant, $f_n$, of a $n {}^3S_1$ quarkonium
state is defined by
\begin{equation}
  \label{cur}
  \langle 0| j^{\mu}_{{\rm em}} |n {}^3S_1\rangle\, = \, f_n M_n
  \epsilon^{\mu} ,
\end{equation}
the leptonic decay width of a $n {}^3S_1$ state reads
\begin{equation}
  \label{vRW}
  \Gamma(nS\to e^+e^-)\,=
                     \,\frac{4\pi}{3}\,\frac{e_Q^2\alpha^2 f_n^2}{M_n}
\end{equation}
where $e_Q$ is the charge of the heavy quark the charmonium state
consists of. Reexpressing (\ref{vRW}) in terms of the non-relativistic
wave function at the origin of the configuration space, one arrives at
the famous van Royen-Weisskopf width \cite{roy67}. The use of the
decay constant instead of the wave function at the origin implies  
that we relate the $\jp\to B\overline B$ (or the $n {}^3S_1\to
B\overline B$) widths to the leptonic width. By that means the
uncertainties in the determination of the wave function at the origin
via the usual van Royen-Weisskopf width cancel to a large
extend. These uncertainties arise from relativistic and QCD
corrections which seem to be large \cite{BKKG} but are not well known
\cite{buc81,eic78}. 

The next uncertainty to be mentioned arises from the choice 
of the $c$-quark mass value. In accordance with calculations of the 
charmonium spectrum within non-relativistic potential approaches 
\cite{buc81} and with a global fit of charmonium parameters
\cite{man95} we take 1.5 GeV as the favoured value. That value 
has, for instance, also been used in a recent analysis of $P$-wave 
charmonium decays into two pions \cite{BKS96}. In spite of this,
little changes of the $m_c$ value cannot be excluded and lead to an
approximate rescaling of the decay widths by the factor $(1.5 \gev/m_c)^{8}$.

The value of $\Lambda_{QCD}$ is also subject to uncertainties. A
change of that value by, say, $\pm 20\,\mev$ which roughly
represents the inaccuracy of our present knowledge of
$\Lambda_{QCD}$ \cite{pdg96}, would alter the theoretical decay 
widths by about $\pm 25\%$. We stress that for any changes of the $m_c$ and 
$\Lambda_{QCD}$ values the ratios of any two decay widths calculated
by us remain approximately unchanged. This assertion does not only
refer to the $\jp \to B_8\overline B_8$ decay widths but it also
applies to the still to be discussed $\jp\to B_{10}\overline B_{10}$
and $\psi'\to B\overline B$ widths.  
 
In addition to the three-gluon contribution ${\mathcal B}_{3g}^{B_8}$
there is a subdominant, although in some cases perhaps sizeable, 
isospin-violating electromagnetic one, ${\mathcal B}_{\rm em}^{B_8}$ 
\cite{cla82}, arising from the graph shown in Fig.\
\ref{fig:graphs}b. This contribution is proportional to the time-like 
magnetic form factor of the baryon. For the proton 
${\mathcal B}_{\rm em}^p$ amounts to about 15 \% (in absolute value) 
of the total ${\mathcal B^p}$ as is estimated from the recent 
measurement of the proton magnetic form factor in the time-like region
\cite{arm93}. Since the relative phase between ${\mathcal B}_{3g}^p$ 
and ${\mathcal B}_{\rm em}^p$ is unknown we cannot simply add the two 
contributions 
\footnote
         {The time-like nucleon form factor is likely not under the regime of
           perturbative QCD in the energy region of interest; it is
           about a factor of 3 larger in absolute value than the form
           factor in the space-like region at $Q^2=\mpsi^2$.
}.
Therefore, we merely can state that, depending on the value of the
relative phase, the electromagnetic contribution (including the
interference term between it and ${\mathcal B}_{3g}^{p}$) to the 
$\jp \to p\overline p$ decay width can be as large as 30 \% or only 
2\% . The comparison of our result with data indicates a relative 
phase close to $\pm \pi/2$ in the $p\overline p$ case.

At this point a remark concerning the $n \overline n$ decay channel
is in order. Since, as said repeatedly, the three-gluon contribution 
respects isospin symmetry any difference between the $p \overline p$ and  
$n \overline n$ decay widths must be due to the electromagnetic 
contribution. From experiment it is known that the widths for 
$\jp \to p \overline p$ and $\jp \to n \overline n$ decays agree 
within the experimental errors \cite{Ant93a} and that the time-like 
form factors of the proton and the neutron are approximately equal 
in modulus at least at $s = 5.4$ GeV$^2$ \cite{Ant93b}. 
Thus, one may conclude that the relative phases 
between the three-gluon and the electromagnetic contributions are 
the same (up to a possible sign) for the proton and the neutron channel. 

The size of the electromagnetic contribution to the hyperon channels 
may be estimated from a recent analysis of the octet baryon form factors
within a diquark model \cite{JKS93}. With the help of a few rather
well determined parameters that model is able to describe a large 
number of exclusive observables. In particular relevant for the
present work is the prediction that, in the space-like region, the
magnetic form factors of the $\Sigma^+$ and $\Sigma^-$ have opposite
signs and are comparable in magnitude to that of the proton.
The form factors of the $\Lambda$ and $\Sigma^0$, on the other hand, 
turn out to be very small. Predictions for the $\Xi^-$ form 
factor are not reported in \cite{JKS93} but that form factor is 
presumably smaller in absolute value than the proton form factor. 
Assuming similar relative magnitudes of the form factors in 
the time-like region, we expect that, in the case of the hyperon 
channels listed in Tab.\ \ref{tab:Gamma}, the three-gluon
contributions should match with the experimental data. This may not be
the case for the $\Sigma^-\overline \Sigma{}^+$ channel; while the
three-gluon contribution is the same as for the 
$\Sigma^0\overline \Sigma{}^0$ channel, the electromagnetic
contribution may be large.

There is also a small contribution to the invariant ${\mathcal B}^{B_8}$ 
from $c \bar c$ annihilations mediated by two gluons and a photon. 
The $gg\gamma$ contribution to ${\mathcal B}^{B_8}$ being proportional 
to the $ggg$ contribution \cite{cla82}, amounts to less than about 
1\% of the latter and is therefore neglected.

Finally, from the measurement of the angular distribution of 
$B_8 \overline B_8$ pairs produced in $e^+ e^- \to \jp \to B_8 \overline B_8$ 
\cite{DM2} one observes small violations of the helicity sum rule: 
The fraction of $p\overline p$ and $\Lambda\overline \Lambda$ pairs 
with equal helicities amounts to about 10\% of the total number of 
pairs with, in particular in the $\Lambda$ case, large errors. For the
other hyperon channels the errors are so large that no conclusion can
be drawn. The small amount of equal helicity pairs is what is to be 
expected if the process is dominated by perturbative QCD: Each of the 
virtual gluons creates a quark and an antiquark with opposite 
helicities. Since our baryon wave functions do not embody any
non-zero orbital angular momentum component the quark helicities sum 
up to the baryon helicity. Hence, baryon and antibaryon are 
produced with opposite helicities. The small amount of 
$B_8\bar B_8$ pairs with the wrong helicity combination observed 
experimentally, while indicating the presence of some soft
contributions, can be considered as a hint that perturbative QCD is 
the dominant dynamical mechanism in the $\jp \to B_8\overline B_8$
decays. One should, however, be aware of these contributions when 
theoretical results for the $\jp \to B_8 \overline B_8$ decays are
compared with experiment. The production of $B_8\overline B_8$ pairs 
with equal helicities can perhaps be explained as an
constituent quark mass and/or baryon mass efffect \cite{cla82,car87}.

%
%%%%%%%%%%%%%%%%%%%%%%%%%%%%%%%%%%%%%%%%%%%%%%%%%%%%%%%%%%%%%%%%%%
\section{$\jp$ decays into decuplet baryons}
%%%%%%%%%%%%%%%%%%%%%%%%%%%%%%%%%%%%%%%%%%%%%%%%%%%%%%%%%%%%%%%%%%
%
\setcounter{equation}{0}
We are now going to apply the modified perturbative approach to the 
$\jp$ decays into the decuplet baryon-antibaryon channels 
$\Delta^{++}\overline{\Delta}{}^{--}$ and 
$\Sigma^{\star-}\overline{\Sigma}{}^{\star+}$
in the same manner as for the octet baryon case. As we said above, within a 
perturbative approach and with wave functions of zero orbital angular 
momentum components in the direction of the baryon momentum, $B_{10}$ 
and $\overline B_{10}$ are produced with opposite helicities. For the 
same reason only helicities $\pm 1/2$ are possible. Thus, as for the 
octet baryon case, the only non-zero helicity amplitudes are the 
${\mathcal M}^{B_{10}}_{\pm\mp\lambda}$. They are fed by only one 
invariant ${\mathcal B}^{B_{10}}$ (or, depending on the definitions of
the covariants, only one linear combination of the invariants) out of the 
five the general covariant decomposition of the $\jp\to B_{10}\overline B_{10}$ 
helicity amplitudes comprises. Within the modified perturbative approach, 
the three-gluon contribution to the invariant ${\mathcal B}^{B_{10}}$ reads
\begin{align}
  {\mathcal B}^{B_{10}}_{3g} =&  \sqrt{\frac{3}{2}}\frac{f_{\psi}}{4}\,
     \int [{\rm d}x] [{\rm d}x'] \int\,\frac{{\rm d}^2{\bf b}_1}{(4\pi)^2}\,
     \frac{{\rm d}^2{\bf b}_3}{(4\pi)^2} \,
     \hat T_H(x,x',{\bf b})\,\exp[-S(x,x',2m_c)] 
     \nonumber \\ 
   \times \, & 
   \quad \hat\Psi^{B_{10}}_{123}(x,{\bf b}) \hat\Psi^{B_{10}}_{123}(x',{\bf b})  
     \, , 
\label{B3gDec} 
\end{align}
for  $B_{10} = \Delta,\Sigma^{\star}$. The three-gluon contribution is 
evaluated with the decuplet wave functions introduced in Sect.\ 3 and 
with the hard scattering amplitude (\ref{THfour2}). Insertion of
${\mathcal B}^{B_{10}}_{3g}$ into (\ref{GamB}) provides the wanted decay 
widths. It still remains to choose plausible values of the parameters 
$a_{(10)}$ and $f_{(10)}$. The fact that we assume the same form for the 
$k_{\perp}$ dependence of the $\Delta$ and the nucleon wave functions, 
and that we use $\phi_{\mathrm AS}$ for the $\Delta$ \da\ which does 
not differ from the actual nucleon \da\ (\ref{phiFIT}) greatly, suggests, 
as a first attempt, the ansatz $a_{(10)}=a_{(8)}\; (=0.75\, \gev^{-1})$ 
and $P^{\Delta}_{3q} = P^{N}_{3q}$. Thereby, it is perhaps plausible
to evaluate the valence quark probability of the nucleon from 
$\phi_{\mathrm AS}$ instead from (\ref{phiFIT}). Doing so we find 
$P^N_{3q}({\mathrm AS})=0.163$ and the requirement 
$P^{\Delta}_{3q} =P^{N}_{3q}({\mathrm AS})$ leads to
$f_{(10)}=0.0163 \gev^2$ which is larger than $f_{(8)}$ by a factor 
of $\sqrt{6}$, i.e., the SU(6) result. The results for the $\jp$ decay
widths into $B_{10}\overline B_{10}$ pairs are presented in 
Tab.\ \ref{tab:Decuplet-DAs}. As can be seen our result for the
$\Sigma^{\star}\overline{\Sigma}{}^{\star}$ channel is too large as 
compared to the data while agreement is achieved for the 
$\Delta \overline\Delta$ case.
%
%%%%%%%%%%%%%%%%%%%%%%%%%%%%%%%%%%%%%%%%%%%%%%%%%%%%%%%%%%%%%%%%%%%%%%%
%    T A B L E   4  :  f_{(10)}, B_n^{Sigma} and J/Psi widths              %
%%%%%%%%%%%%%%%%%%%%%%%%%%%%%%%%%%%%%%%%%%%%%%%%%%%%%%%%%%%%%%%%%%%%%%%
%
\begin{table}
  \begin{tabular}{|c||c|r|r|r|r|r||c|c|} \hline
    \rule{0cm}{8mm}${{\dst a_{(10)}} \atop {\dst [\mathrm{GeV}^{-1}]}}
               \atop {\rule{0cm}{0mm}}$ & 
    \rule{0cm}{8mm}${{\dst f_{(10)}} \atop {\dst [\mathrm{GeV}^2]}}
               \atop {\rule{0cm}{0mm}}$ & $B_1\;\;$ &
    $B_2\;\;$ & $B_3\;\;$ & $B_4\;\;$ & $B_5\;\;$ &
    $\Gamma_{3g}^{\Delta \overline \Delta}$ [eV] & 
    $\Gamma_{3g}^{\Sigma^\star \overline {\Sigma}{}^\star}$ [eV]  \\ \hline\hline
    0.75  & 0.0163 
            & -0.494 &  0.165 & -0.203 & -1.013 & 0.058 
            & 105 & 66.1 \\ \hline
    0.80  & 0.0143 
            & -0.547 &  0.182 & -0.216 & -1.081 & 0.062 
            & 82.6 & 51.8 \\ \hline
    0.85  & 0.0127 
            & -0.601 &  0.200 & -0.229 & -1.142 & 0.065
            & 65.1 & 40.8  \\ \hline\hline
    & \multicolumn{6}{|c||}{ $\Gamma_{\rm exp}$ [eV]
               \hspace*{0.5cm} \cite{pdg96}}
     & $96 \pm 26$ & $45 \pm 6$ \\ \hline     
  \end{tabular}
  \caption[]{$f_{(10)}$, the expansion coefficients $B_n$ for the 
    $\Sigma^{\star}$ \da\ and results for the three-gluon contribution
    to the $\jp\to\Delta^{++}\overline{\Delta}{}^{--},\;
           \Sigma^{\star-}\overline{\Sigma}{}^{\star+}$ decay widths 
    for various values of the transverse size parameter $a_{(10)}$
    ($m_s=350\, \mev$). $f_{(10)}$ is fixed by the requirement 
    $P_{3q}^{\Delta}= 0.163$. 
\label{tab:Decuplet-DAs}}
\end{table}

Bearing in mind that the $\Delta$ is in completely symmetric flavour 
and spin states and that the Pauli principle effectively induces an 
additional repulsive interquark force, one may expect a larger
radius, and hence a larger value of $a_{(10)}$, for the $\Delta$
than for the nucleon. Therefore, we also try values for $a_{(10)}$ 
slightly larger than that for $a_{B_8}$ and fix in each case 
$f_{(10)}$ from the requirement $P^{\Delta}_{3q}=0.163$ as before. 
In Tab.\ \ref{tab:Decuplet-DAs} we list results obtained with the 
values 0.80 and 0.85 GeV$^{-1}$ for $a_{(10)}$. For both these values of 
$a_{(10)}$ satisfactory agreement with experiment is found. The 
uncertainties of the theoretical results are the same as for the 
octet baryons. In particular, one has to consider the possibility of large
electromagnetic contributions (see Fig.\ \ref{fig:graphs}b).

%
%%%%%%%%%%%%%%%%%%%%%%%%%%%%%%%%%%%%%%%%%%%%%%%%%%%%%%%%%%%%%%%%%%
\section{Decays of the $\psi'$ and other quarkonia}
%%%%%%%%%%%%%%%%%%%%%%%%%%%%%%%%%%%%%%%%%%%%%%%%%%%%%%%%%%%%%%%%%%
%
\setcounter{equation}{0}
The extension of our approach to baryonic decays of the $\psi'$
($=\psi(2\,^3S_1)$) is now a simple matter. It is however important to
realize that, in contrast to other authors \cite{BrL81}, we evaluate
the three-gluon contribution with the $c$-quark mass and not with the
charmonium mass. This is, as we said, legitimate in a non-relativistic
treatment of the charmonia. Hence, in order to get the
$\psi'$ widths in our approach we have not to rescale the
corresponding $\jp$ widths by $(\mpsi/M_{\psi'})^8$ but rather by 
\begin{equation}
  \label{sca}
  \Gamma(\psi'\to B\overline B)\,=\,
         \frac{\rho_{\mathrm p.s.}(m_{B}/M_{\psi'})}
                                {\rho_{\mathrm p.s.}(m_{B}/\mpsi)}\;
         \frac{\Gamma(\psi'\to e^+ e^-)}{\Gamma(\jp \to e^+ e^-)}\;
                   \Gamma(\jp\to B\overline B)     
\end{equation}
which holds for both, octet and decuplet baryons. The charmonium decay
constants are replaced by the leptonic decay widths by means of 
formula (\ref{vRW}). 

As an immediate examination of our approach one may apply the scaling
relation (\ref{sca}) directly to the experimental data (see Tabs.\
\ref{tab:Gamma}, \ref{tab:Decuplet-DAs}, \ref{tab:2S}). Considering
the uncertainties due to the electromagnetic contributions, which in
one or the other case may be large, (\ref{sca}) works quite
well in particular for the octet baryons. For the decuplet baryons, on
the other hand, it seems that the suppression of the $\psi'$ decay
widths is a slightly underestimated, although the large experimental errors
prevent any definite conclusion at present. One may suspect the
neglect of the decuplet baryon masses to be responsible for that
possible imperfection. In any case, an additional strong suppression, 
as provided by $(\mpsi/M_{\psi'})^8$ \cite{BrL81}, is in clear conflict
with the data \cite{BESC}. This observation supports our attempt of 
using the $c$-quark mass in the calculation of the decay amplitudes 
rather than the mass of the charmonium state in question.
%
%%%%%%%%%%%%%%%%%%%%%%%%%%%%%%%%%%%%%%%%%%%%%%%%%%%%%%%%%%%%%%%%%%%%%%%
%    T A B L E   6  :  psi' decays                                    %
%%%%%%%%%%%%%%%%%%%%%%%%%%%%%%%%%%%%%%%%%%%%%%%%%%%%%%%%%%%%%%%%%%%%%%%
%
\begin{table}
  \begin{center}
  \begin{tabular}{|c||c|c|c|c|c|c|} \hline
    \rule{0cm}{8mm} channel & $p\overline p\;\;$ &
    $\Sigma^0\overline{\Sigma}{}^0\;\;$ & $\Lambda\overline \Lambda\;\;$ & 
    $\Xi^-\overline{\Xi}{}^+\;\;$ & $\Delta^{++}\overline{\Delta}{}^{--}\;\;$ &
    $\Sigma^{*-}\overline{\Sigma}{}^{*+}$ \\ \hline\hline
    $\Gamma_{3g}$  & 76.8 & 55.0 & 54.6 & 33.9 & 32.1 & 24.4\\ \hline  
    $\Gamma_{\rm exp}$ \cite{BESC} & $76\pm 14$ 
            & $26\pm 14$ &  $58\pm 12$ & $23\pm 9$ & $25\pm 8$ & $16\pm 8$ \\ 
    {\phantom{$\Gamma_{exp}$}} \cite{pdg96}  & $53\pm 15$ 
            & &   &  &  & \\ \hline
  \end{tabular}
  \end{center}
  \caption[]{The three-gluon contribution to the $\psi'\to B\overline B$ 
    decay widths (in [eV]) computed through (\ref{sca}) from the set 3
    $\jp$ widths (see Tab.\ \ref{tab:Gamma}). $a_{(10)}=0.85\,\gev^{-1}$.
\label{tab:2S}}
\end{table}

Results for baryonic decay widths of the $\psi'$, evaluated through
(\ref{sca}) from the set 3 $\jp$ widths, are listed in Tab.\
\ref{tab:2S} where also recent experimental results of the BES 
collaboration \cite{BESC} are quoted. The data are still preliminary.
The agreement between theoretical results (with
$a_{(10)}=0.85\,\gev^{-1}$ in the decuplet baryon case) and experiment
is generally good although our results seem to be a bit too large for the
$\Sigma\overline \Sigma$ and $\Sigma^*\overline \Sigma^*$ channels. 
For a discussion of uncertainties we refer to Sect.\ 3.

Computation of the $\psi(3\,^3S_1)$ decay widths are difficult within our
approach. The relativistic corrections are presumably larger since the
$\psi(3\,^3S_1)$ mass is above the threshold for open charm production. The
$B\overline B$ decay widths are likely to be very small and it is
hardly conceivable that they will be measured. Therefore, we refrain
from estimating these decay widths.

Results for bottonium decays, on the other hand, can safely be
calculated within our approach. The hard scale, provided by the 
$b$-quark mass, is larger than in the charmonium case and relativistic
corrections are smaller. But, as it turns out, the predicted decay
widths for the baryonic channels are also very small. Approximately, i.e.\
ignoring the fact that the $k_{\perp}$-dependent suppression of the
three-gluon contribution is a little bit different in the two cases, we
find the following rescaling formula
\begin{equation}
  \label{Ypsi}
  \Gamma(\Upsilon\to B\overline B)\,=\,4\,
         \frac{\rho_{\mathrm p.s.}(m_{B}/M_{\Upsilon})}
                                {\rho_{\mathrm p.s.}(m_{B}/\mpsi)}\;
         \frac{\Gamma(\Upsilon\to e^+ e^-)}{\Gamma(\jp \to e^+ e^-)}\;
          \left (\frac{m_c}{m_b}\right)^8 \Gamma(\jp\to B\overline B)     
 \end{equation}
Using $m_b=4.5\,\gev$ we obtain, for instance, a value of $0.03$ eV
for the $\Upsilon\to p\overline p$ decay width which value
corresponds to a branching ratio of $0.5\times 10^{-6}$ well below the
experimental upper bound \cite{pdg96}. The decay widths for the other 
$B\overline B$ channels are even smaller.

%
%%%%%%%%%%%%%%%%%%%%%%%%%%%%%%%%%%%%%%%%%%%%%%%%%%%%%%%%%%%%%%%%%%
\section{Summary}
%%%%%%%%%%%%%%%%%%%%%%%%%%%%%%%%%%%%%%%%%%%%%%%%%%%%%%%%%%%%%%%%%%
%
\setcounter{equation}{0}
In this investigation, we applied the modified perturbative approach to
the decays of $\jp$ and $\psi'$ into baryon-antibaryon
pairs. We demonstrated that, on the basis of plausible baryon wave 
functions for which $\su3$ symmetry is only mildly broken by quark mass
effects and for which even SU(6) symmetry, in sharp contrast to the 
QCD sum rule based wave functions, approximately holds, 
the experimental data for octet and decuplet baryon channels are quite
well reproduced by the phase space corrected three-gluon
contributions. The perturbative contributions to the decay widths are 
calculated self-consistently in the sense that the bulk of a
perturbative contribution is accumulated in regions of reasonably
small values of $\al$. Besides the form of the wave functions used by 
us our analysis differs from previous ones in the following points:\\
i) The use of the modified perturbative approach allows to take into
account the running $\al$ and the evolution of the wave functions
properly, in contrast to the usual leading twist analysis. The
virtualities of the internal $c$ quarks and gluons can be chosen as 
the arguments of $\al$. Hence, $\al$ reflects the characteristic 
scale of the process under question. The running coupling constant is 
therefore not a quasi-free parameter that can, within a certain range,
be chosen arbitrarily. Since the decay widths are proportional to 
$\al^6$, a large factor of uncertainty is therefore hidden in the standard
perturbative analysis. \\
ii) The hard scattering amplitude is computed with the $c$-quark mass
instead of the charmonium mass. The use of the
$c$-quark mass is consistent with the non-relativistic treatment of
the charmonium state and with the perturbative approach. Hence the
$\psi'$ - $\jp$ scaling relation (\ref{sca}) holds in our approach.
That scaling relation is nicely confirmed by the data in contrast to
the usual $(\mpsi/M_{\psi'})^8$ scaling. 

Although our results for the decay widths agree with the data, we are
aware of a number of uncertainties in our calculation. Above all, we
mention as a source of uncertainty the value of the $c$-quark
mass. The value we have chosen (1.5 \gev) is consistent with other
constraints on that mass. Another uncertainty arises from the 
electromagnetic contribution which, at least in the $p\overline p$ 
case, can be large.  In virtue of the unknown relative phase between 
the electromagnetic and the three-gluon contribution, the first cannot 
be taken into account properly. We emphasize that most of the
uncertainties cancel in ratios of widths to a large extend.

It should be noted that we have not found hints at substantial
contributions from the $c\overline c g$ Fock state, i.e., colour octet
contributions \cite{BKS96}, neither in the baryonic $\jp$ nor in the
$\psi'$ decays.

An interesting class of $\jp$ decays are the $B_8 \overline B_{10}$
channels. While the three-gluon contributions to the 
$p\overline \Delta^-$ and $\Sigma^{*0}\overline \Lambda$ channels are 
strictly zero \cite{koe87} they are non-zero - although small - for other
$B_8 \overline B_{10}$ channels, since our wave functions exhibit only
a mild breaking of $\su3$ symmetry. Experimentally, only upper
bounds are known for the first two channels \cite{pdg96} saying
that these decays are indeed suppressed by at least an order of
magnitude as compared to, say, the $p\overline p$ or the
$\Lambda\overline \Lambda$ channels. The experimental widths for the
other $B_8 \overline B_{10}$ channels are surprisingly large
\cite{pdg96}. Thus, in accordance with \cite{koe87}, we expect as the
dominant $\su3$ breaking mechanism for these reactions a sizeable
electromagnetic contribution.  

Finally, we have to mention that there are a few exclusive charmonium
decays which cannot be described within the standard or the modified 
perturbative approach. Thus, the relatively large branching ratio of
the process $\jp\to\rho\pi$ observed experimentally, for instance,
indicates a substantial violation of hadronic helicity conservation
while only mild violations are observed in the baryonic $\jp$ decays (see
the discussion in Sect. 3). For a discussion of this puzzle and a
possible solution of it by means of a hypothetical glueball see
\cite{bro87}.

Acknowledgement. We thank S.J.\ Brodsky, J.G.\ K\"orner, D.\ Lichtenberg and
N.G.\ Stefanis for useful discussions and suggestions. We are also
grateful to Y.\ Zhu for comments on the BES data. 

%
%%%%%%%%%%%%%%%%%%%%%%%%%%%%%%%%%%%%%%%%%%%%%%%%%%%%%%%%%%%%%%%%%%%%%%%
%                        R E F E R E N C E S                          %     
%%%%%%%%%%%%%%%%%%%%%%%%%%%%%%%%%%%%%%%%%%%%%%%%%%%%%%%%%%%%%%%%%%%%%%%
%

%

\begin{thebibliography}{99}
\bibitem{BrL81} S.J.\ Brodsky and G.P.\ Lepage,
                \Journal{\PRD}{24}{2848}{1981}.
\bibitem{COZ89b}V.L.\ Chernyak, A.A.\ Ogloblin and I.R.\ Zhitnitsky,
                \Journal{\ZPC}{42}{583}{1989}.
\bibitem{BeS92} N.G.\ Stefanis and  M.\ Bergmann,
                \Journal{\PRD}{47}{3685}{1993};
                \Journal{\PLB}{304}{24}{1994}.
\bibitem{app75} Th.\ Appelquist and H.D.\ Politzer,
                \Journal{\PRL}{34}{43}{1975}.
\bibitem{wubo}  J.\ Bolz, R.\ Jakob, P.\ Kroll, M.\ Bergmann,
                and N.G.\ Stefanis,\\
                \Journal{\ZPC}{66}{267}{1995}.
\bibitem{BoK96} J.\ Bolz and P.\ Kroll, 
                \Journal{\ZPA}{356}{327}{1996}. 
\bibitem{rad82} V.A.\ Nesterenko and A.V.\ Radyushkin,
                \Journal{\em Yad.\ Fiz.}{39}{1287}{1984}. 
\bibitem{BoSLi} J.\ Botts and G.\ Sterman,
                \Journal{\NPB}{325}{62}{1989}; \\
                H.-N.\ Li and G.\ Sterman,
                \Journal{\NPB}{381}{129}{1992}.
\bibitem{BESC}  Y.\ Zhu for the BES coll., talk presented at the 
                XXVIII Int.\ Conf.\ on High Energy Physics, 
                25-31 July 1996, Warsaw, Poland.
\bibitem{Sot}   M.G.\ Sotiropoulos and G.\ Sterman,
                \Journal{\NPB}{425}{489}{1994}.
\bibitem{Dzi88} Z.\ Dziembowski, \Journal{\PRD}{37}{768}{1988}.
\bibitem{BrL80} G.P.\ Lepage and S.J.\ Brodsky,
                \Journal{\PRD}{22}{2157}{1980}.
\bibitem{ste94} N.G.\ Stefanis, 
                \Journal{\em Acta Phys.\ Pol.\ B}{25}{1777}{1994}.
\bibitem{ste89} N.G.\ Stefanis, \Journal{\PRD}{40}{2305}{1989}; 
                \Journal{D}{44}{1616(E)}{1991}.
\bibitem{BHL83} S.J.\ Brodsky, T.\ Huang and G.P.\ Lepage, 
                Banff Summer Institute, 
                Particles and Fields 2, p.\ 143,
                A.Z.\ Capri, A.N.\ Kamal (eds.) (1983).
\bibitem{COZ89a}V.L.\ Chernyak, A.A.\ Ogloblin and I.R.\ Zhitnitsky,
                \Journal{\ZPC}{42}{569}{1989}.
\bibitem{far89} G.R.\ Farrar, H.\ Zhang, A.A.\ Ogloblin and I.R.\
                Zhitnitsky, \Journal{\NPB}{311}{585}{1988/89}.
\bibitem{BKS96} J.\ Bolz, P.\ Kroll and G.S.\ Schuler,
                \Journal{\PLB}{392}{198}{1997}. 
\bibitem{bar81} R.\ Barbieri, R.\ Gatto and E.\ Remiddi,
                \Journal{\PLB}{106}{497}{1981}.
\bibitem{bod95} G.T.\ Bodwin, E.\ Braaten and G.P.\ Lepage,
                \Journal{\PRD}{51}{1125}{1995}. 
\bibitem{buc81} W.\ Buchm\"uller and S.-H.\ Tye,
                \Journal{\PRD}{24}{132}{1981}.
\bibitem{DJK95} M.\ Dahm, R.\ Jakob and P.\ Kroll,
                \Journal{\ZPC}{68}{595}{1995}.
\bibitem{CoS81} J.C.\ Collins and D.E.\ Soper,
                \Journal{\NPB}{193}{381}{1981}.
\bibitem{pdg96} R.M.\ Barnett et al., 
                \Journal{\PRD}{54}{1}{1996} 
                (Review of Particle Properties).
\bibitem{eat84} M.W.\ Eaton et al., MARK2 collaboration,
                \Journal{\PRD}{29}{804}{1984}.
\bibitem{DM2}   D.\ Pallin et al., DM2 collaboration,
                \Journal{\NPB}{292}{653}{1987}.
\bibitem{hen87} P.\ Henrard et al., DM2 collaboration,
                \Journal{\NPB}{292}{670}{1987}.
\bibitem{Ant93a} A.\ Antonelli et al., FENICE collaboration,
                \Journal{\PLB}{301}{317}{1993}.
\bibitem{roy67} R.\ van Royen and V.F.\ Weisskopf,
                 \Journal{\NCA}{50}{617}{1967}.
\bibitem{BKKG}  R.\ Barbieri, R.\ K\"ogerler, Z.\ Kunszt and R.\ Gatto,
                \Journal{\NPB}{105}{125}{1976}.
\bibitem{eic78} E.\ Eichten, K.\ Gottfried, T.\ Kinoshita, K.D.\ Lane
                and T.-M.\ Yan, \Journal{\PRD}{17}{3090}{1978} and D
                {\bf 21}, 203 (1980).
\bibitem{man95} M.L.\ Mangano and A.\ Petrelli, 
                \Journal{\PLB}{352}{445}{1995}.
\bibitem{cla82} M.\ Claudson, S.L.\ Glashow and M.B.\ Wise, 
                \Journal{\PRD}{25}{1345}{1982}.
\bibitem{arm93}  T.A.\ Armstrong et al., E760 collaboration, 
                \Journal{\PRL}{70}{1212}{1993}. 
\bibitem{Ant93b} A.\ Antonelli et al., FENICE collaboration,
                \Journal{\PLB}{313}{283}{1993}.
\bibitem{JKS93} R.\ Jakob, P.\ Kroll, M.\ Sch\"urmann and W.\ Schweiger,
                \Journal{\ZPA}{347}{109}{1993}.
\bibitem{car87} C.\ Carimalo, 
                \Journal{\em Int.\ J.\ Mod.\ Phys.\ A}{2}{249}{1987}.
\bibitem{koe87} J.G.\ K\"orner,
                \Journal{\ZPC}{33}{529}{1987}.
\bibitem{bro87} S.J.\ Brodsky, G.P.\ Lepage and S.F.\ Tuan, 
                \Journal{\PRL}{59}{621}{1987}.
\end{thebibliography}
\end{document}